\documentclass[a4paper,fleqn,usenatbib]{mnras}
\pdfoutput=1

\usepackage{newtxtext,newtxmath}

\usepackage[T1]{fontenc}
\usepackage{ae,aecompl}

\usepackage[lofdepth,lotdepth]{subfig}

\usepackage{graphicx}	
\usepackage{amsmath}	
\usepackage{amssymb}	

\title[Tidal dynamos in radiative stars]{Magnetic fields driven by tidal mixing in radiative stars}

\author[Vidal et al.]{
J. Vidal$^{1}$\thanks{jeremie.vidal@univ-grenoble-alpes.fr},
D. C\'ebron$^{1}$,
N. Schaeffer$^{1}$
and R. Hollerbach$^{2}$
\\
$^{1}$Universit\'e  Grenoble Alpes, CNRS, ISTerre, 38000 Grenoble, France\\
$^{2}$Department of Applied Mathematics, University of Leeds, Leeds LS2 9JT , UK\\
}

\date{Accepted XXX. Received YYY; in original form ZZZ}

\pubyear{2018}

\begin{document}
\label{firstpage}
\pagerange{\pageref{firstpage}--\pageref{lastpage}}
\maketitle

\begin{abstract}
	Stellar magnetism plays an important role in stellar evolution theory.
    Approximatively 10~\% of observed main sequence (MS) and pre-main-sequence (PMS) radiative stars exhibit surface magnetic fields above the detection limit, raising the question of their origin.
    These stars host outer radiative envelopes, which are stably stratified. Therefore, they are assumed to be motionless in standard models of stellar structure and evolution.
	We focus on rapidly rotating, radiative stars which may be prone to the tidal instability, due to an orbital companion.
	Using direct numerical simulations in a sphere, we study the interplay between a stable stratification and the tidal instability, and assess its dynamo capability.
	We show that the tidal instability is triggered regardless of the strength of the stratification (Brunt-V\"ais\"al\"a frequency).
    Furthermore, the tidal instability can lead to both mixing and self-induced magnetic fields in stably stratified layers (provided that the Brunt-V\"ais\"al\"a frequency does not exceed the stellar spin rate in the simulations too much).
    The application to stars suggests that the resulting magnetic fields could be observable at the stellar surfaces. Indeed, we expect magnetic field strengths up to several Gauss.
	Consequently, tidally driven dynamos should be considered as a (complementary) dynamo mechanism, possibly operating in radiative MS and PMS stars hosting orbital companions.
	In particular, tidally driven dynamos may explain the observed magnetism of tidally deformed and rapidly rotating Vega-like stars.
\end{abstract}



\section{Introduction}
	\subsection{Stellar magnetism}
Stellar magnetic fields were first discovered in the Sun \citep{hale1908probable} and in the chemically peculiar Ap star 78 Virginis \citep{babcock1947zeeman}. 
Stellar magnetism sparks growing interest, since it provides additional data to infer the dynamical processes occurring in stellar interiors.
On one hand, it has been known for decades that magnetic fields are common in solar-like low mass stars, in which magnetic fields have complex surface structures and time variabilities. 
Since the pioneering works of \citet{larmor1919could,parker1955hydromagnetic,roberts1968thermal,busse1970thermal}, many works in stellar magnetism have considered magnetic fields driven by thermo-chemical convection. Indeed, it is widely accepted that stellar magnetic fields originate from motions within the convective envelope, generating dynamo action \citep{parker1979cosmical}.
Convectively driven dynamo action is supported by magnetohydrodynamic numerical simulations of both stellar and planetary fluid interiors \citep[e.g.][]{glatzmaiers1995three,brun2004global,schaeffer2017geodynamo,strugarek2017reconciling}.
Furthermore, reduced mean-field or flux-transport models can be tuned to reproduce magnetic cycles as observed for the Sun \citep[e.g.][]{jouve2007role,jouve2010exploring,charbonneau2014solar} or solar-like stars \citep[e.g.][]{jouve2010exploring}.

On the other hand, the magnetism of hot Ap/Bp stars, a group of intermediate-mass A/B stars showing strong chemical peculiarities, with outer radiative layers (i.e. stably stratified in density), is different from the magnetism of cool solar-like stars.
Indeed, they display global dipolar fields, with typical amplitudes ranging from 300 Gauss \citep{auriere2007weak} to thousands of Gauss, and seem remarkably stable over observational time \citep{donati2009magnetic}.
Recently, magnetic fields with Gauss-level amplitudes have been detected in several stars \citep{blazere2016discovery,blazere2016detection}, e.g. in Vega \citep{lignieres2009first,petit2010rapid} and in Sirius A \citep{petit2011detection}. They form another class of magnetic stars defining the Vega-like stellar magnetism. Hence, there is a strong dichotomy, or magnetic desert, between strong and ultra-weak magnetic fields among hot stars \citep{lignieres2013dichotomy}. More generally, astronomical observations show that between 5~\% and 10~\% of main-sequence (MS) (e.g. Ap/Bp) and pre-main-sequence (PMS) (e.g. Herbig Ae/Be) stars exhibit surface magnetic fields \citep{donati2009magnetic,braithwaite2017magnetic,mathys2017ap}.

It is commonly accepted that stars form from a fully convective low-mass core, which grows through accretion during the proto-stellar phase \citep{palla1992evolution,behrend2001formation}. 
However, hot stars undergo important changes in their interior structures before reaching the main sequence.
Stellar models indicate that after the initial fully convective phase, a radiative core forms and grows in the whole star. This suggests that sun-like dynamo action does not occur in hot stars with thick outer radiative envelopes. However, in very massive stars, an innermost convective core may develop. Hence, the magnetic desert may result from the large variability of mechanisms generating magnetic fields in hot stars. 

	\subsection{Proposed mechanisms in hot stars}
The origin of stellar magnetism in hot stars remains elusive and debated \citep{neiner2014origin}. The observed fields are often presumed to be fossil fields \citep{borra1982magnetic,braithwaite2004fossil}, which were shaped during the stellar formation phase \citep{power2008properties} and might survive into later stages of stellar evolution. The observed strong dipolar fields of Ap/Bp stars are stable over time \citep{donati2009magnetic}, which is compatible with fossil fields. However, it seems difficult for rapidly rotating stars to reach stable magnetic equilibrium \citep{braithwaite2012weak}. Similarly, the fossil field model does not seem to predict the observed small-scale and weak fields of Vega-like stars. It has been proposed that their magnetic field is at equilibrium but undergo a dynamical evolution before reaching an equilibrium state \citep{braithwaite2012weak}.
Moreover, the fossil field origin has also been questioned for the magnetic field of pre-main-sequence (PMS) Herbig Ae/Be stars, which are expected to be the precursors of magnetic Ap/Bp stars on the PMS phase  \citep{alecian2012high}. 
However, the recently observed dramatic change of the surface magnetic field of HD 190073 \citep{alecian2013dramatic}, which possibly hosts a small inner convective core, could result from interactions with a dynamo field generated in the convective core.

Hence, dynamo action could also take place in the small inner convective cores of some hot stars \citep{stello2016prevalence}. It is argued that surface fields could be due to the emergence of magnetic field blobs produced by a powerful convective dynamo in the innermost core \citep{parker1975generation,charbonneau2001magnetic}. However, the time required for this dynamo field to reach the stellar surface may be longer than the lifetime of the star \citep{moss1989origin,macgregor2003magnetic}, unless very thin magnetic tubes could be generated. 
Moreover, in radiative interiors only magnetic fields much stronger than the equipartition value in the innermost convective core are able to be carried out to the stellar surface, which challenges the core-dynamo model \citep{macdonald2004magnetic}.
Interactions between a fossil field and a core dynamo are also possible, leading to a super-equipartition state in the convective core \citep{featherstone2009effects}.

In early-type O and B stars, a sub-surface convective layer may exist and a dynamo could develop in this layer \citep{cantiello2011magnetic}. This mechanism produces magnetic fields of strength between 5 and 50~G, rather small scale and time-dependent, while the observed fields are mainly dipolar, stable over time and of much stronger amplitude. In intermediate-mass stars (smaller than $8 M_\odot$), such as Vega and Sirius, sub-surface convective layers are also expected \citep{cantiello2011magnetic}, although being of different physical nature. Nevertheless, the dynamo action in such thin layers is unlikely to sustain magnetic fields of large-enough length scales to be detectable \citep{kochukhov2013detectability}.

Another hypothesis relies on a dynamo action in the radiative envelope.
Indeed, differential rotation can trigger various instabilities which lead to dynamo action, as shown by self-consistent numerical simulations \citep{macdonald2004magnetic,guervilly2010numerical,arlt2011amplification,marcotte2016dynamo}.
Several instabilities are likely to occur in stellar interiors \citep{spruit1999differential}.
Dynamo cycles (of the $\alpha\Omega$-type), based on flux-tube instabilities \citep[e.g.][]{ferriz1994dynamo,zhang2003three}, the magneto-rotational instability \citep{balbus1991powerful,mizerski2012connection} or the pinch-type Tayler instability \citep{tayler1973adiabatic,markey1973adiabatic,pitts1985adiabatic} have been proposed. In stably stratified envelopes, a pinch-type instability is expected to be the first to occur \citep{spruit1999differential}. 
Thus, recent theoretical and experimental works \citep{gellert2011helicity,seilmayer2012experimental,weber2015tayler} focused on the Tayler instability in fluids with low magnetic Prandtl number, but yielded contradictory results. 
The dynamo capability of the Tayler instability in radiative envelopes was considered by \citet{spruit2002dynamo,braithwaite2006differential}. This mechanism is conceptually similar to the one driven by the magneto-rotational instability \citep[e.g.][]{jouve2015three}. An initial axisymmetric poloidal seed field is transformed by the $\Omega$ effect into an axisymmetric toroidal field. 
Then, a magnetic instability in the toroidal field develops to generate non-axisymmetric field components. To close the dynamo loop, a regeneration of either an axisymmetric toroidal \citep{spruit2002dynamo} or poloidal field \citep{braithwaite2006differential} is invoked. \citet{braithwaite2006differential} conducted numerical simulations, which seem to validate the dynamo mechanism in stellar stratified interiors. This dynamo mechanism has been criticised by \citet{zahn2007magnetic}. They used numerical simulations that did not lead to dynamo action. However, these simulations considered high magnetic diffusivity, yielding a differential rotation in these simulations below the threshold for dynamo action \citep{braithwaite2017magnetic}. Later, \citet{arlt2011magnetic,szklarski2013nonlinear} observed dynamo action in numerical simulations.
Finally, \citet{jouve2015three} found that the magneto-rotational instability seems favored at the expense of the Tayler instability in differentially rotating, incompressible stars. 

Undoubtedly, clarifying the relevance of these dynamo mechanisms in more realistic models of stably stratified stars deserves future work.
Observational tests should play an essential role. In particular, a correlation between the stellar rotation and the magnetic field properties should exist \citep[e.g.][]{potter2012stellar}, but this is not observed \citep{hubrig2006discovery,mathys2017ap}.
Then, in all scenarios based on differential rotation, an energy source for that differential rotation needs to be identified.
Indeed, the toroidal field is produced by shearing the poloidal field and it draws its energy from the differential rotation. As a result, this mechanism could only operate as long as a differential rotation exists.
However, magnetohydrodynamic effects tend to weaken the initial differential rotation, which may be provided by the stellar contraction occurring during the PMS phase,
through dissipative processes \citep{arlt2003differential,jouve2015three}. 
Ultimately, the latter effects weaken the energy source of the dynamo action. Strong field strengths at the stellar surface are also expected to warrant a uniformly rotating radiative envelope \citep{spruit1999differential}, for instance in B3.5V star HD 43317 \citep{buysschaert2017magnetic}.

Tidal forcing is another possible mechanism in radiative stars, as long as stars host non-synchronised orbital companions. Indeed, tidally deformed fluid bodies are prone to the tidal instability \citep[e.g.][]{kerswell2002elliptical,cebron2013elliptical,barker2016non,vidal2017inviscid}. The latter is a hydrodynamic instability of elliptical streamlines that excites inertial waves through parametric resonance. The nonlinear outcome of the tidal instability could lead to space-filling turbulence \citep[e.g.][]{barker2013nona,barker2013nonb,barker2016nonb,le2017inertial}. 
It has been proposed that the tidal instability is of significant importance for tidal dissipation in binary systems \citep{rieutord2004evolution,le2010tidal} and for angular momentum transport in accretion discs \citep{goodman1993local}. 
The dynamo capability of the tidal instability has been confirmed by numerical simulations \citep{barker2013nonb,cebron2014tidally}. Apart from dynamo action in hot stars, it has also been shown that a Hot Jupiter companion is responsible for the stellar activity enhancement of low-mass HD 179949 star \citep{fares2012magnetic}. The role of the close-in massive planet in the short activity cycle of the star $\tau$ Bootis has also been suggested \citep{fares2009magnetic}.
Finally, tides might also lead to a resonant excitation of helical oscillations driven by the Tayler instability, suggesting a possible planetary synchronisation of the solar dynamo \citep{stefani2016synchronized}.

	\subsection{Motivations}
On one hand, the hydrodynamic nonlinear regime of the tidal instability has been studied in unstratified fluids by \citet{cebron2010systematic,barker2013nona,barker2016nonb,grannan2016tidally}.
The tidal instability can induce a magnetic field \citep{lacaze2006magnetic,herreman2010elliptical}, paving the way to dynamos as suggested by \citet{mizerski2012mean}. Its dynamo capability has been proved by local \citep{barker2013nonb} and global numerical simulations \citep{cebron2014tidally}.
On the other hand, the nonlinear regime of the tidal instability in stably stratified fluids has been studied by \citet{cebron2010tidal}, but only for a very limited range of parameters. 
It remains unclear how the the tidal instability is modified in stably stratified layers.
Consequently, the main purpose of this numerical study is to investigate the nonlinear outcome of the tidal instability in stably stratified fluids and then to assess its dynamo capability. 

Numerical simulations of the tidal instability are difficult to carry out. The parameter space of stellar interiors is impossible to simulate with the available computational resources. To simulate more realistic configurations we may use local models. Local simulations of the tidal instability in periodic boxes \citep[e.g.][]{barker2013nona,barker2013nonb,le2017inertial} indeed give quantitative predictions in good agreement with global simulations \citep{cebron2010systematic,cebron2014tidally,barker2016non,barker2016nonb} and laboratory observations \citep{le2010tidal,grannan2016tidally}. 
However, it is unclear whether possible small-scale dynamos obtained with local models could lead to large-scale magnetic fields in stellar interiors.

Here, we use global numerical simulations to study the tidal instability and its coupling to a magnetic field.
In such simulations, the internal magnetic field matches a potential field outside the tidally deformed domain -- such as a triaxial ellipsoid. This is a source of great mathematical complexity in non-spherical geometries \citep[e.g.][]{wu2009dynamo}.
Existing numerical codes capable of handling ellipsoidal boundaries -- such as codes based on finite elements \citep{cebron2010tidal,cebron2012magnetohydrodynamic}, spectral finite elements \citep{favier2015generation,barker2016nonb} or finite volumes \citep{vantieghem2015latitudinal} -- approximate this magnetic boundary condition at the cost of both low accuracy and slow execution.
However, high performance is crucial to try to reach the low viscosity limit relevant for astrophysical bodies.
We choose to perform proof-of-principle numerical simulations in a spherical container. By considering a sphere we benefit from the efficiency and accuracy of spectral codes relying on spherical harmonics \citep{schaeffer2013efficient,matsui2016performance}.
We extend the method proposed by \citet{cebron2014tidally} to handle stratification. We assume that the fluid is subjected to a non-conservative body force sustaining an analytically designed tidally driven flow, valid in spherical geometry and satisfying the various constraints (including the viscous boundary condition).
This flow is then prescribed in the code, and we consider the departure from the basic state.

The paper is organised as follows. 
In section \ref{sec:model}, we present the mathematical and numerical formulations of the problem.
Numerical results are presented and discussed in depth in section \ref{sec:results}. 
Then, we extrapolate our results to stellar interiors in section \ref{sec:astrophysics}.
Section \ref{sec:ccl} ends the paper with a discussion and perspectives. 

\section{Description of the problem}
\label{sec:model}
\subsection{Governing equations}
We model tides in a rotating fluid sphere of radius $R_*$. We consider a Newtonian fluid of uniform kinematic viscosity $\nu$, thermal diffusivity $\kappa$ and magnetic diffusivity $\eta = 1/(\mu_0 \sigma_e)$, where $\sigma_e$ is the electrical conductivity and $\mu_0$ the magnetic permeability of free space. The fluid is rotating with the spin spin angular velocity $\Omega_s \boldsymbol{\widehat{z}}$ along the vertical axis.
We consider the variations of density only in the buoyancy force, using the Boussinesq approximation \citep{spiegel1960boussinesq}. 
The density $\rho$ is given by the non-barotropic equation of state
\begin{equation}
	\rho = \rho_* \left [ 1 - \alpha (T-T_*) \right ],
	\label{eq:EoS}
\end{equation}
with $\alpha$ the coefficient of thermal expansion, $(\rho_*, T_*)$ typical density and temperature and $T$ the departure of the temperature field from the adiabatic temperature profile. 
In the Boussinesq framework the fluid is stratified under the gravity field $\boldsymbol{g} = - \nabla \Phi_0$, with $\Phi_0$ a prescribed gravitational potential.
We choose $R_*$ as unit of length, $\Omega_s^{-1}$ as unit of time, $\Omega_s^2 R_* / (\alpha g_0)$ as unit of temperature $T$, where $g_0$ is the gravitational acceleration at the stellar surface, and $R_* \Omega_s \sqrt{\mu_0 \rho_*}$ as unit of magnetic field $\boldsymbol{B}$. We introduce the dimensionless Ekman number $Ek=\nu/(\Omega_s R_*^2)$, the Prandtl number $Pr = \nu/\kappa$ and the magnetic Prandtl number $Pm=\nu/\eta$. 
To quantify the stratification we introduce the dimensionless (local) Brunt-V\"ais\"al\"a frequency $N (\boldsymbol{r})$ defined by \citep{friedlander1982internala}
\begin{equation}
	N^2 (\boldsymbol{r}) = -\alpha \boldsymbol{g} \boldsymbol{\cdot} \nabla T,
	\label{eq:BruntVaisala}
\end{equation} 
The fluid is stably stratified if $N^2 > 0$. 

We work in spherical coordinates $(r, \theta, \phi$). 
We expand in the inertial frame the velocity field and the temperature as perturbations $(\boldsymbol{u}, \Theta, \boldsymbol{B})$ around a steady tidally driven basic state $(\boldsymbol{U}_0, T_0, \boldsymbol{0})$.
In the inertial frame, the dimensionless non-ideal, nonlinear magnetohydrodynamic equations are
\begin{subequations}
	\label{eq:goveqn}
	\begin{align}
		\frac{\partial \boldsymbol{u}}{\partial t} &= -( \boldsymbol{u} \boldsymbol{\cdot} \boldsymbol{\nabla} ) \, \boldsymbol{U}_0 - ( \boldsymbol{U}_0 \boldsymbol{\cdot} \boldsymbol{\nabla} ) \, \boldsymbol{u} - ( \boldsymbol{u} \boldsymbol{\cdot} \boldsymbol{\nabla} ) \, \boldsymbol{u} - \nabla p + Ek \, \nabla^2 \boldsymbol{u} \nonumber \\
		{} &- \Theta \, \boldsymbol{g} + (\boldsymbol{\nabla} \times \boldsymbol{B}) \times \boldsymbol{B}, \label{eq:goveqnU} \\
		\frac{\partial \Theta}{\partial t} &= - ( \boldsymbol{U}_0 \boldsymbol{\cdot} \nabla ) \, \Theta - ( \boldsymbol{u} \boldsymbol{\cdot} \nabla ) \, T_0 - ( \boldsymbol{u} \boldsymbol{\cdot} \nabla) \, \Theta + \frac{Ek}{Pr} \nabla^2 \Theta, \label{eq:goveqnT}\\
		\frac{\partial \boldsymbol{B}}{\partial t} &=  \boldsymbol{\nabla} \times (\boldsymbol{U}_0 \times \boldsymbol{B}) + \boldsymbol{\nabla} \times (\boldsymbol{u} \times \boldsymbol{B}) + \frac{Ek}{Pm} \nabla^2 \boldsymbol{B}, \label{eq:goveqnB} \\
		\boldsymbol{\nabla} \boldsymbol{\cdot} \boldsymbol{u} &= 0, \ \, \ \boldsymbol{\nabla} \boldsymbol{\cdot} \boldsymbol{B} = 0, \label{eq:goveqndivUB}
	\end{align}
\end{subequations}
with $p$ the modified pressure, ensuring the incompressibility of the dynamics. 
For hydrodynamic computations, the Lorentz force $(\boldsymbol{\nabla} \times \boldsymbol{B}) \times \boldsymbol{B}$ is removed. 
Equations (\ref{eq:goveqn}) are supplemented with appropriate boundary conditions. The velocity field satisfies the stress-free boundary condition
\begin{equation}
	\boldsymbol{u} \boldsymbol{\cdot} \boldsymbol{n} = 0, \ \, \ \boldsymbol{n} \times \left [ \boldsymbol{n} \boldsymbol{\cdot} (\boldsymbol{\nabla}\boldsymbol{u} + (\boldsymbol{\nabla} \boldsymbol{u}) ^T) \right ] = \boldsymbol{0},
	\label{eq:stressfree}
\end{equation}
where $\boldsymbol{n}$ is the unit radial vector. 
Following \citet{cebron2014tidally} we impose a zero-angular momentum for $\boldsymbol{u}$.
We also assume a fixed temperature $\Theta = 0$ at the boundary.
Finally, the external region ($r>1$) is assumed to be electrically insulating. Thus, the magnetic field matches a potential field at the boundary. 

The governing equations (\ref{eq:goveqn}) are solved with the open-source parallel XSHELLS code \citep[e.g.][]{schaeffer2017geodynamo}. It has been validated against standard benchmarks \citep{marti2014,matsui2016performance}.
It uses second  order finite differences in radius and pseudo-spectral spherical harmonic expansion, handled efficiently by the free SHTns library \citep{schaeffer2013efficient}.
The time-stepping scheme is of second order in time, and treats the diffusive terms implicitly, while the nonlinear and Coriolis terms are handled explicitly.
For this study, we have extended the XSHELLS code to handle arbitrary basic state fields.

All simulations have been performed at $Ek=10^{-4}, Pr=1$ with various $N_0/\Omega_s$ and $Pm$.
The spatial discretisation uses $N_r=224$ radial points, $l_{\max}=128$ spherical harmonic degrees and $m_{\max}=100$ azimuthal wave numbers.
We made sure that our simulations are numerically converged by varying the spatial resolution.

\begin{figure}
	\centering
	\begin{tabular}{cc}
	\subfloat[Ellipticity]{
		\includegraphics[width=0.2\textwidth]{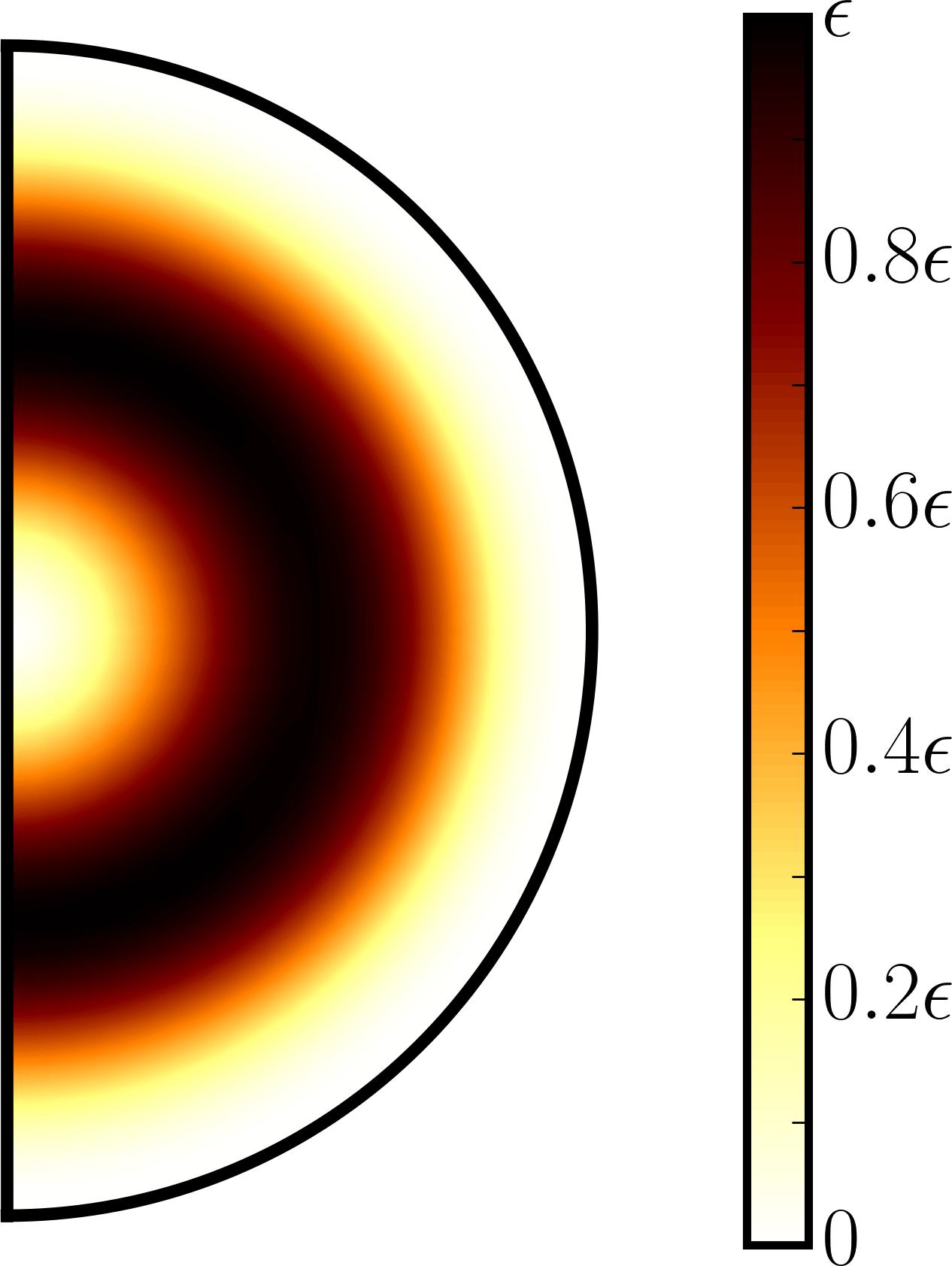}
		}
		& 
	\subfloat[Brunt-V\"ais\"al\"a frequency]{
		\includegraphics[width=0.215\textwidth]{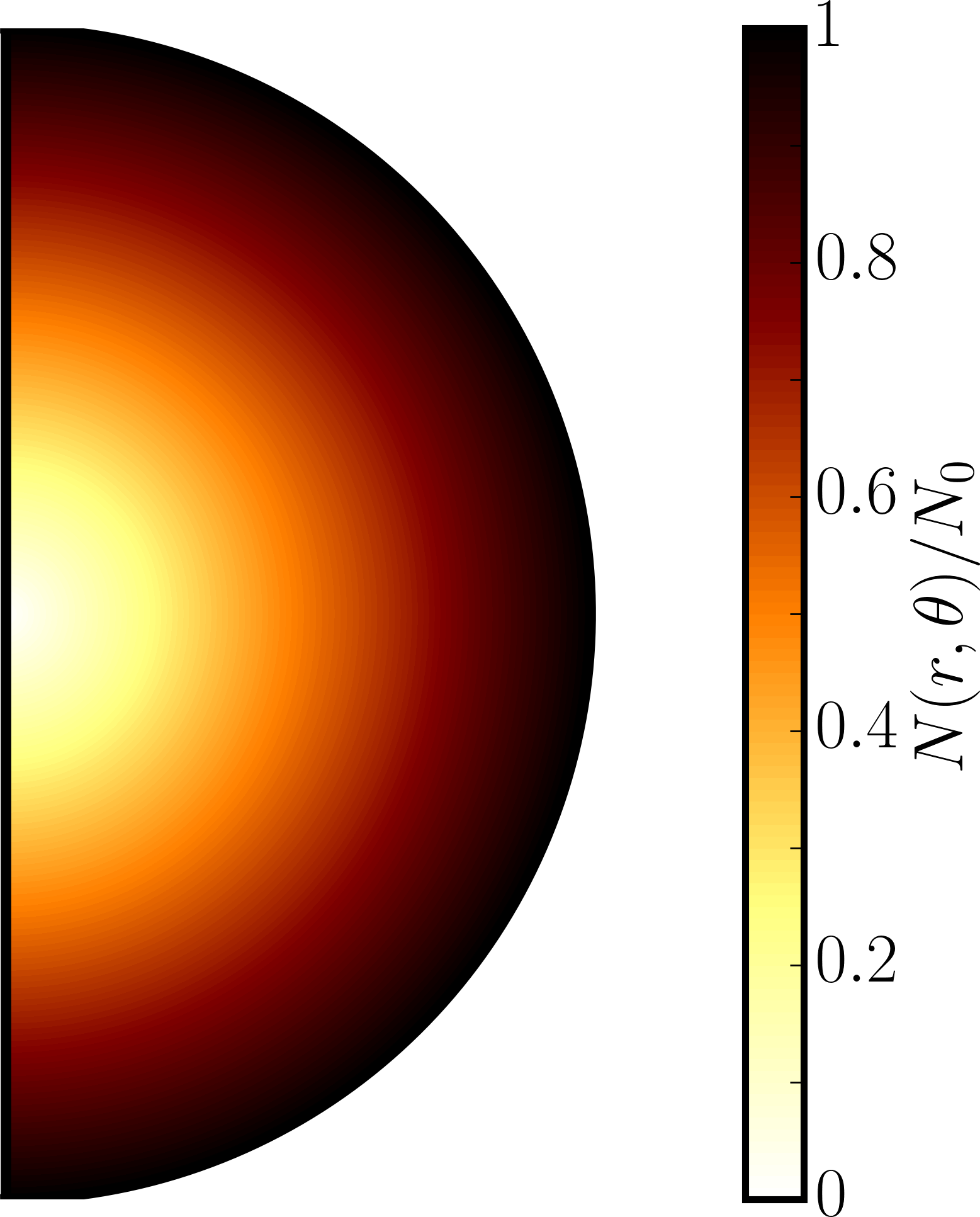}
		}
		\\
	\end{tabular}
	\subfloat[Ellipticity and Brunt-V\"ais\"al\"a frequency versus radius]{
		\includegraphics[width=0.4\textwidth]{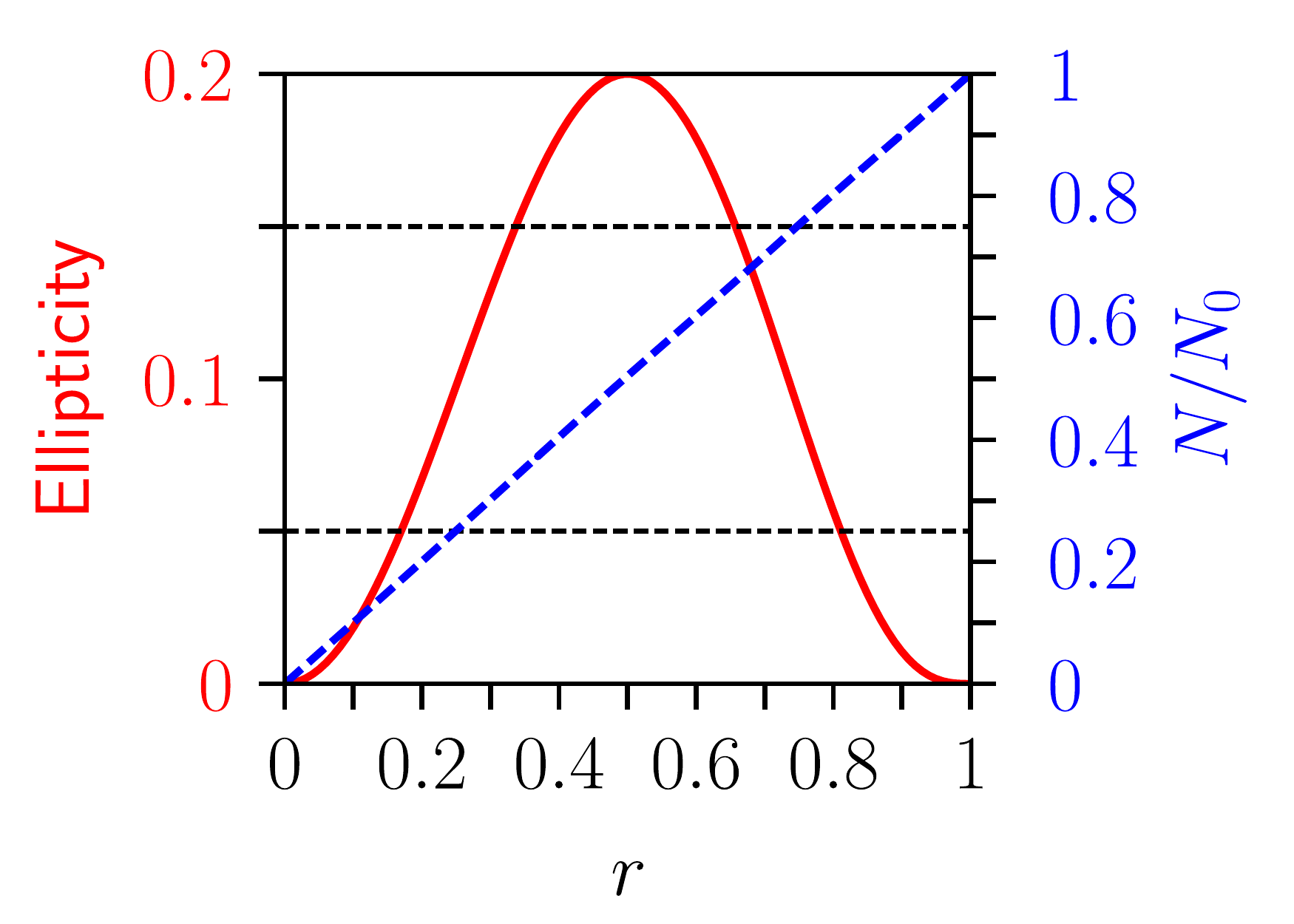}
		}
	\caption{(a) Ellipticity $\epsilon f(r,\theta)$ in a meridional plane, computated at $\epsilon=0.2$. (b) Normalised Brunt-V\"ais\"al\"a frequency of the basic state $N(r, \theta)/N_0$ in the meridional plane $\phi=0$. (c) Equatorial ellipticity  $\epsilon f(r,\pi/2)$ and $N(r, \theta)/N_0$ are shown as solid and dashed lines respectively. Horizontal dashed lines represent the critical ellipticity $\epsilon_c=0.05$ and $\epsilon_c=0.15$, for $N_0/\Omega_s \ll 1$ and $N_0/\Omega_s \simeq 2$ respectively.}
	\label{fig:Ellipticity_Meridional}
\end{figure}

\subsection{Tidal basic state}
The disturbing tidal potential perturbs the spin solid-body rotation to generate a flow with elliptical streamlines, known as the equilibrium tide \citep[e.g.][]{zahn1966marees,remus2012equilibrium}. A difficulty is to numerically establish the equilibrium tide in spherical geometry. Following \citet{favier2014non}, we can impose numerically a non-zero radial flow, or similarly decompose the flow into non-wavelike and wavelike parts \citep{ogilvie2005wave,rieutord2010viscous,ogilvie2013tides,lin2017tidal}.
However, the relevance of these methods are elusive for dynamo computations, because the fluid  suddenly becomes insulating when crossing the spherical boundary.

We assume that the fluid is subjected to a non-conservative body force $\boldsymbol{f}$ and heat source term $\mathcal{Q}$.
They aim at deforming the axisymmetry (mimicking tidal effects), yielding the basic flow $\boldsymbol{U}_0$ and the basic temperature $T_0$. As in the non-wavelike decomposition \citep[e.g.][]{rieutord2010viscous}, the body force $\boldsymbol{f}$ is vortical \citep{cebron2014tidally}, i.e. $\boldsymbol{\nabla} \times \boldsymbol{f} \neq \boldsymbol{0}$. This is a necessary condition to deform the the circular streamlines of the solid-body rotation into elliptical ones in incompressible fluids.

Instead of directly prescribing $\boldsymbol{f}$ \citep{cebron2014tidally} and $\mathcal{Q}$ in the governing equations (\ref{eq:goveqnBFU}), we prescribe an analytical basic flow $\boldsymbol{U}_0$ and temperature $T_0$. Indeed, imposing $(\boldsymbol{U}_0, T_0)$ is computationally more efficient, because we solve the departure from the basic state. Moreover the imposed tidally driven basic state satisfies the various boundary constraints (including the viscous boundary condition).
The imposed analytical basic state $(\boldsymbol{U}_0, T_0)$, is an exact steady solution of the primitive equations in the inertial frame
\begin{subequations}
	\label{eq:goveqnBF}
	\begin{align}
		(\boldsymbol{U}_0 \boldsymbol{\cdot} \boldsymbol{\nabla}) \boldsymbol{U}_0 &= - T_0 \, \boldsymbol{g} - \nabla P_0 + Ek \, \nabla^2 \boldsymbol{U}_0 + \boldsymbol{f}, \label{eq:goveqnBFU} \\
		( \boldsymbol{U}_0 \boldsymbol{\cdot} \nabla ) \, T_0 &= \frac{Ek}{Pr} \nabla^2 T_0 + \mathcal{Q}, \label{eq:goveqnBFT} \\
		\boldsymbol{\nabla} \boldsymbol{\cdot} \boldsymbol{U}_0 &= 0. \label{eq:goveqnBFdivU}
	\end{align}
\end{subequations}
The body force $\boldsymbol{f}$ and the heat source term $\mathcal{Q}$ can be analytically computed from equations (\ref{eq:goveqnBF}). Their mathematical expressions is rather lengthy and so they are not provided here. 

The basic state depends solely on a stream function $\Psi_0$ as follows.
The disturbing tidal potential generates an elliptical flow of azimuthal wave number $m=2$, superimposed on the spin solid body rotation ($m=0$). For simplicity we consider a dimensionless basic flow of the form 
\begin{equation}
	\boldsymbol{U}_0 (\boldsymbol{r}) = \boldsymbol{\nabla} \times [ \Psi_0 (\boldsymbol{r}) \boldsymbol{\widehat{z}} ],
\end{equation}
where $\Psi_0 (\boldsymbol{r})$ is a stream function given by
\begin{equation}
	\Psi_0 (\boldsymbol{r}) = -\frac{r^2}{2} + \epsilon f(r,\theta) \cos(2\phi),
	\label{eq:PsiU0}
\end{equation}
with $\epsilon$ the maximum equatorial ellipticity and $f(r, \theta) \leq 1$ the local ellipticity profile. The effective ellipticity is $\beta (r, \theta) = \epsilon f(r, \theta)$. The latter profile is built to ensure that the basic flow $\boldsymbol{U}_0$ satisfies the stress-free boundary condition (\ref{eq:stressfree}). It is also constrained by a regularity condition at the centre \citep{lewis1990physical}. After little algebra it reads
\begin{equation}
	f(r,\theta) = \frac{256}{9} r^2 \left (\frac{1}{3} - r^2 + r^4 - \frac{1}{3}r^6 \right ) \frac{r^2 \sin^2 \theta}{2}.
	\label{eq:betar}
\end{equation}
The basic flow $\boldsymbol{U}_0$ satisfies the stress-free boundary condition (\ref{eq:stressfree}). It is an approximation of the equilibrium tide \citep{zahn1966marees,remus2012equilibrium}. 

Then, we choose a background temperature profile of the form $T_0 = N_0^2/\Omega_s^2 \, \Phi_0$, where $N_0$ is the dimensional Brunt-V\"ais\"al\"a frequency at the outer boundary ($N_0^2 \geq 0$). 
It has a fixed temperature at the boundary and cancels out the baroclinic instability, as a result of $\nabla T_0 \times \boldsymbol{g} = \boldsymbol{0}$. Thus, we ensure a barotropic basic state. We further assume a linear dependence between the imposed gravitational potential $\Phi_0$ and the stream function, i.e. $\Phi_0 = - \Psi_0$. Therefore, isotherms in the basic state coincide with streamlines. With this choice the imposed gravitational potential is constant at the outer spherical boundary ($r=1$). 

Finally, the tidally driven basic state (\ref{eq:PsiU0}) does not take into account the rotation of the tidal ellipticity due to the companion's orbital motion. Indeed, the rotation of the tidal strain does not modify the underlying physical mechanism of the tidal instability \citep{le2010tidal,cebron2014tidally}. In the non-rotating orbital case, the zero angular momentum condition imposed for $\boldsymbol{u}$ is in agreement with the conservation of the angular momentum of the star.

Our basic state is illustrated in figure \ref{fig:Ellipticity_Meridional} for a given set of parameters. The effective tidal ellipticity equals its maximum value $\epsilon$ at $r=0.5$ and decreases towards the centre and the outer boundary where it vanishes.
As a consequence, azimuthal averages of $T_0$, $\boldsymbol{g}$ and of the background Brunt-V\"ais\"al\"a frequency almost vary linearly in radius, as observed in figure \ref{fig:Ellipticity_Meridional} (b). Hence, our basic stratification is almost spherically symmetric, which is expected for rotating stars.

\section{Numerical results}
\label{sec:results}
	\subsection{Hydrodynamic regime}
	\label{sec:hydro}
\begin{figure}
	\centering
	\subfloat[Linear growth]{
		\includegraphics[width=0.45\textwidth]{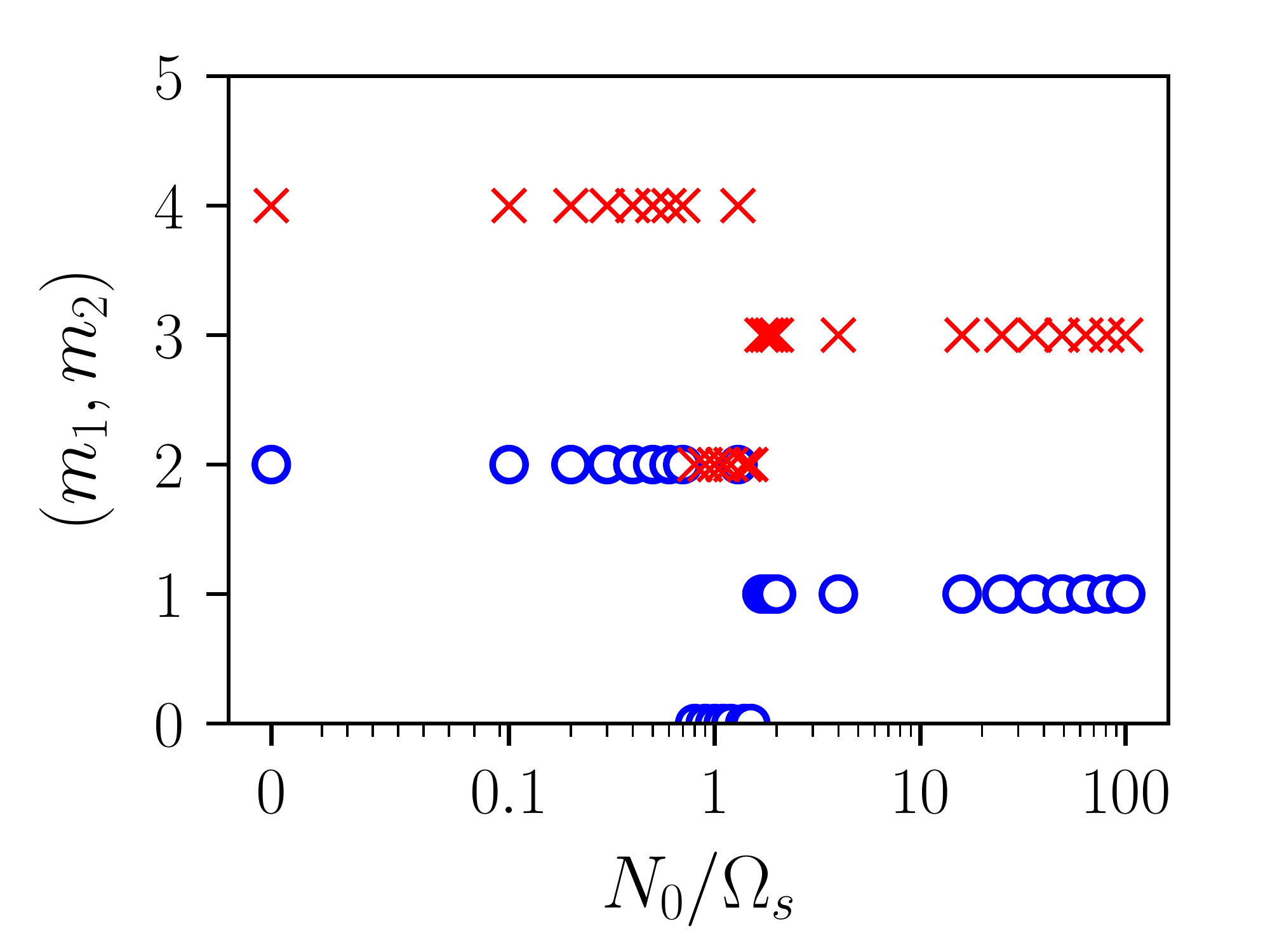}
		}
		\\
	\subfloat[Nonlinear saturation]{
		\includegraphics[width=0.45\textwidth]{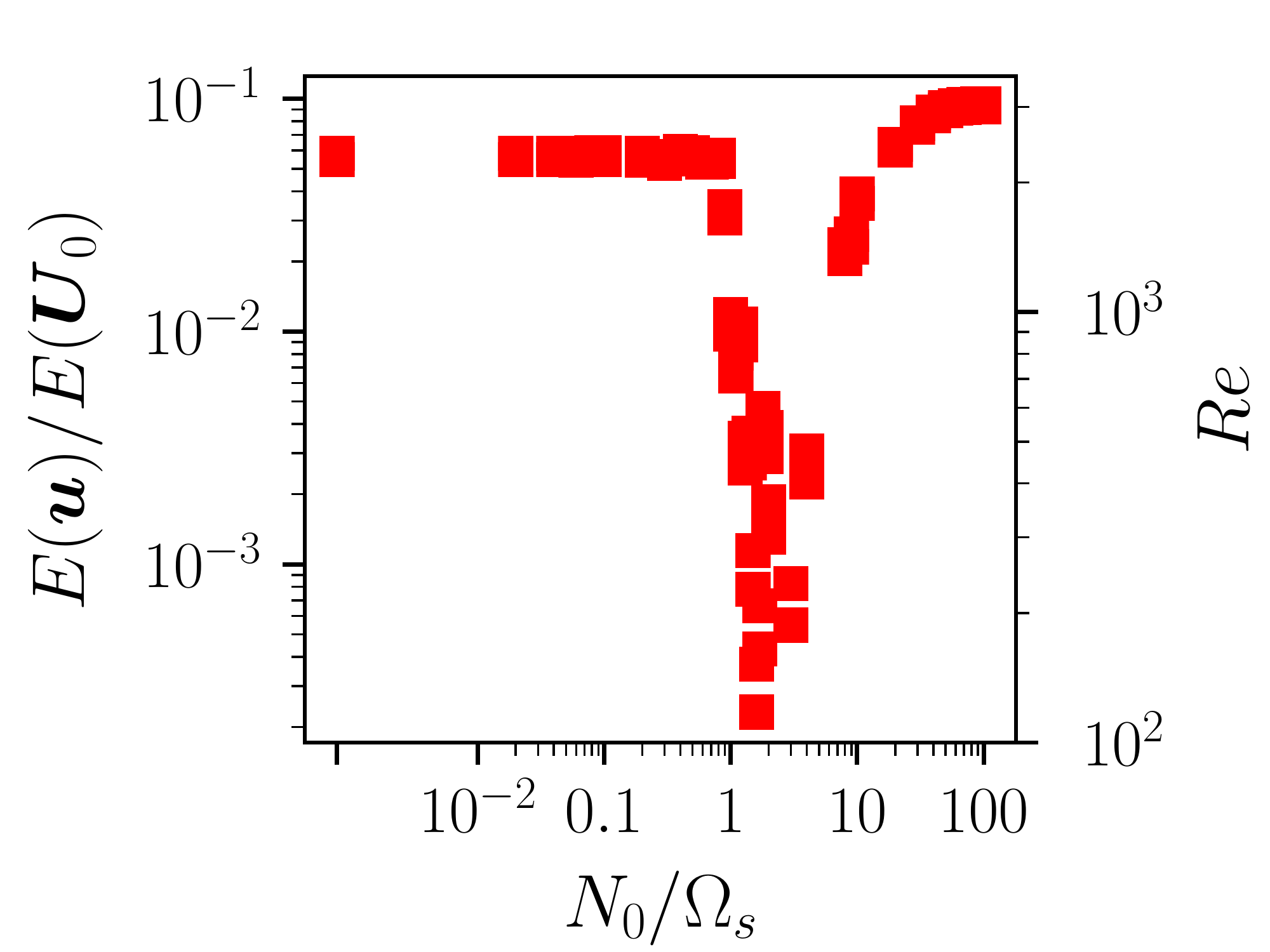}
		}
	\caption{Survey of hydrodynamic simulation of the tidal instability at $Ek=10^{-4}, Pr=1$ and $\epsilon=0.2$ for varying $N_0/\Omega_s$. (a) Pair of the most rapidly growing wave numbers $m_1$ (circles) and $m_2$ (crosses) excited in the exponential growth. (b) Volume average kinetic energy of the perturbation $E(\boldsymbol{u})$  normalised by the kinetic energy of the basic flow $E(\boldsymbol{U}_0)$ and Reynolds number $Re$.}
	\label{Fig_TDEIm_eps02}
\end{figure}

The magnetic field is kept at zero in equations (\ref{eq:goveqn}) to study purely hydrodynamic instabilities of the equilibrium tide $\boldsymbol{U}_0$. When the maximum tidal ellipticity $\epsilon$ is greater than a critical value ($\epsilon_c = 0.054$ in the neutral case $N_0=0$), the basic flow $\boldsymbol{U}_0$ is unstable.
The perturbation flow $\boldsymbol{u}$ grows exponentially and then saturates non-linearly.
Perturbation velocities $\boldsymbol{u}$ of larger amplitudes are obtained for larger $\epsilon$, but we also want to keep $\epsilon$ small enough for the basic state to remain close to a solid body rotation.
Consequently, we choose $\epsilon=0.2 (\simeq 4 \epsilon_c)$ to survey the parameter space in the following.

The nature of the hydrodynamic instability is revealed in the linear growth phase. If the perturbation flow satisfies the global resonance condition \citep{kerswell2002elliptical}
\begin{equation}
	|m_1 - m_2| = 2,
	\label{eq:TDEI_Resonancem}
\end{equation}
where $(m_1, m_2)$ is the azimuthal wave number pair of the inertial modes (i.e. the eigenmodes of a rotating cavity) resonating with the tidal basic flow ($m=2$), then the instability is a tidal instability. 
In figure \ref{Fig_TDEIm_eps02} (a) we show the most energetic wave number pair $(m_1, m_2)$ excited in the exponential growth as function of $N_0/\Omega_s$. All pairs satisfy the condition (\ref{eq:TDEI_Resonancem}), hence a tidal instability is always excited in the explored range of stratification ($0 \leq N_0/\Omega_s \leq 100$). 
It is an equatorially symmetric flow, appearing first at radius $r=0.5$ where the ellipticity is maximum (see figure \ref{fig:Ellipticity_Meridional}), which then spreads out in the bulk.
When $N_0/\Omega_s \lesssim 1$, the pair $(2,4)$ is excited and the typical growth rate is $\sigma/\epsilon \simeq 10^{-1}$ irrespective of the value of $N_0$. It yields typical time scales for the instability to grow between 30~ky to 3~My for typical stellar interiors with $\epsilon \in [10^{-8}, 10^{-6}]$ and a one day spin period. 
The flow oscillates at the angular frequency $\omega \leq 2$, suggesting that the parametric resonance involves inertial modes.
When $1 \lesssim N_0/\Omega_s \leq 2$, we observe different pairs of unstable modes and the growth rates of the tidal instability are lower. In this range, the typical frequencies of inertial modes and internal gravity modes are similar. As a result of the interplay between the two effects of same order of magnitude, a complex pattern of unstable modes is expected. The most unstable pair $(2,4)$ at $N_0=0$ is first replaced by the pair $(0,2)$ when $0.8\leq N_0 \leq 1.5$ and then by the pair $(1,3)$.
When $N_0/\Omega_s \geq 2$, the buoyancy force becomes of primary importance and the stratification is then expected to be always stabilizing \citep{miyazaki1993elliptical}. However, we observe that the tidal instability is not inhibited.
Furthermore, the hydrodynamic growth rates are slightly enhanced by a large stratification, with $\sigma/\epsilon \simeq 5.10^{-1}$. It yields dimensional time scales for the instability to growth of order 5 ky to 0.5 My for typical stellar interiors, with $\epsilon \in [10^{-8}, 10^{-6}]$ and a one day spin period.

\begin{figure}
	\centering
	\subfloat[$N_0/\Omega_s=0.5$]{
		\includegraphics[width=0.35\textwidth]{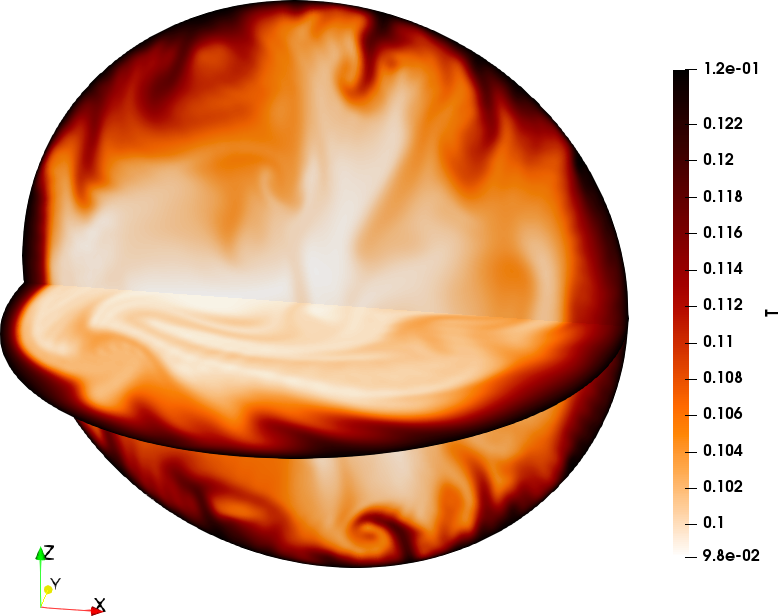}
		}
		\\
	\subfloat[$N_0/\Omega_s=1$]{
		\includegraphics[width=0.35\textwidth]{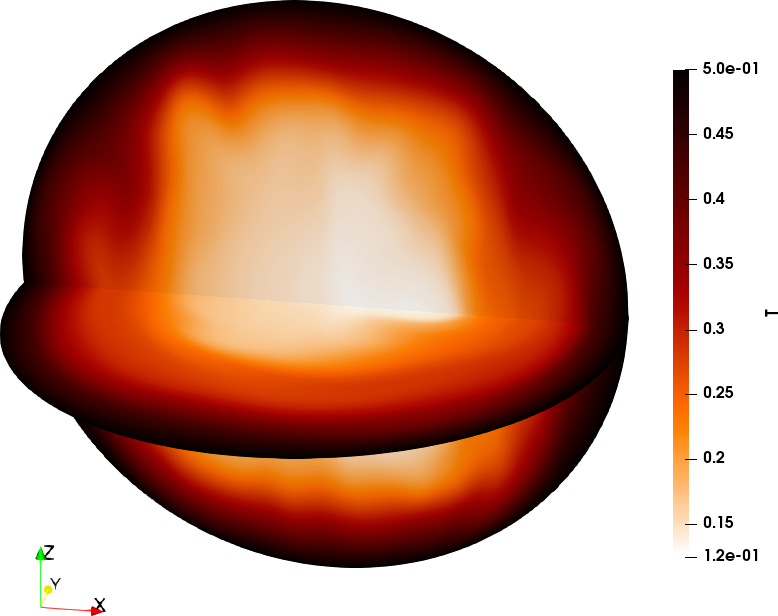}
		}
		\\
	\subfloat[$N_0/\Omega_s=10$]{
		\includegraphics[width=0.35\textwidth]{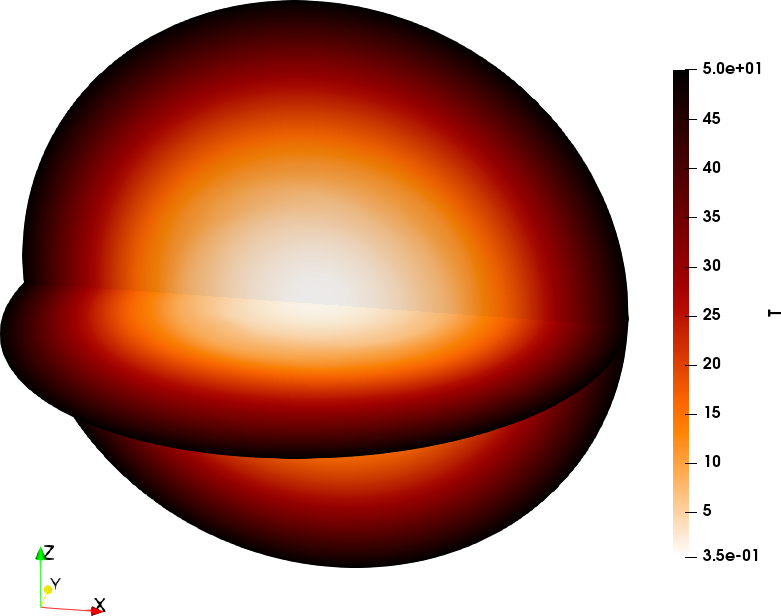}
		}
	\caption{Three-dimensional views of the total temperature $T=T_0 + \Theta$ in the nonlinear regime of the tidal instability. Surfaces of constant $T$ are shown in the equatorial plane and in a meridional plane. Movies are provided in the supplementary materials. Simulations at $Ek=10^{-4}, Pr=1$ and $\epsilon=0.2$}
	\label{Fig_Ttot}
\end{figure}
Finally, the observed pairs $(m_1,m_2)$ depend on the diffusion in our simulations. In asymptotic regime of low diffusion ($Ek \to 0$), we expect the excitation of a wider range of pairs $(m_1,m_2)$, possibly with large azimuthal numbers, leading to wave turbulence \citep{le2017inertial}.

To quantify the nonlinear outcome of the tidal instability, we compute in figure \ref{Fig_TDEIm_eps02} (b) the kinetic energy of the perturbation
\begin{equation}
	E(\boldsymbol{u}) = \int_V \frac{|\boldsymbol{u}|^2 }{2} \mathrm{d} V,
\end{equation}
(with $V = 4 \pi/3$ the dimensionless volume of the sphere), as a function of $N_0/\Omega_s$. We also introduce the Reynolds number $Re = Ro/Ek$ with $Ro= \sqrt{E(\boldsymbol{u})/E(\boldsymbol{U}_0)}$ the Rossby number and $E(\boldsymbol{U}_0)$ the kinetic energy of the global rotation. Three regimes are observed in the simulations. Illustrative three-dimensional snapshots of the total temperature field $T = T_0 + \Theta$ in these regimes are shown in figure \ref{Fig_Ttot}.
When $0 \leq N_0/\Omega_s \lesssim 1$ the tidal instability flow is immune to the stable stratification, as in the linear growth phase.
The instability is almost four times critical in this range ($\epsilon/\epsilon_c \simeq 3.7 $) and the typical Reynolds number is $Re=2000$.
The flow has a kinetic energy representing about 5~\% of the kinetic energy of the global rotation, consistent with the expected dimensional amplitude $\epsilon \, \Omega_s R_*$ in the neutral ($N_0=0$) case \citep{barker2013nona,grannan2016tidally,barker2016nonb}. 
In figure \ref{Fig_Ttot} (a) the stratification seems to be well mixed and eroded in the bulk (compare with fig.1b), below a thermal boundary layer (due to our thermal boundary condition). We note that the fluid is no longer barotropic, since the instantaneous isolines of $T$ do not coincide with the isopotentials anymore.

\begin{figure}
	\centering
	\includegraphics[width=0.45\textwidth]{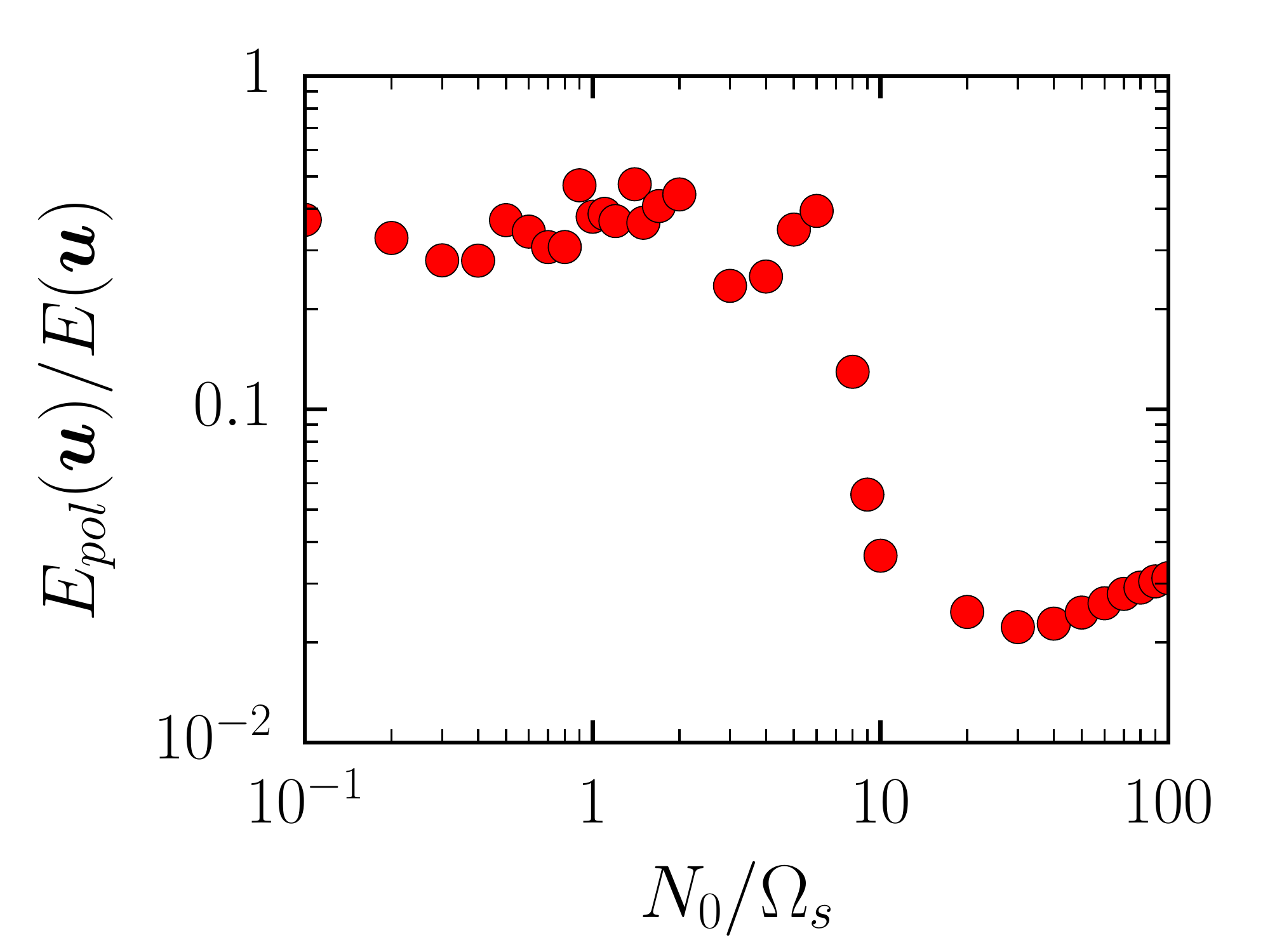}
	\caption{Instantaneous fraction of poloidal to total kinetic energy $E_{pol}(\boldsymbol{u})/E(\boldsymbol{u})$, denoted $F_{pol}$, as a function of $N_0/\Omega_s$. Simulations at $Ek=10^{-4}, Pr=1$ and $\epsilon=0.2$}
	\label{Fig_Fpol}
\end{figure}

When $1 \lesssim N_0/\Omega_s \leq 2$, we observe a collapse of the kinetic energy. For these stratifications the interplay between inertial waves and internal waves reduces the saturation amplitude of the tidal instability. As a consequence, we observe also a reduction in the mixing in figure \ref{Fig_Ttot} (b).
The collapse when $1 \lesssim N_0/\Omega_s \leq 2$ is due to a variation of $\epsilon_c$ there, likely due to diffusion effects (see appendix \ref{sec:append_weakening}).
This effect is not expected in radiative stellar interiors, characterised by much lower diffusivities.  
Finally, when $N_0/\Omega_s \geq 2$, the strong stratifications do not prevent the tidal instability. Instead the instability has an even larger amplitude, with a typical Reynolds number $Re=3000$ and a kinetic energy representing still about 5~\% of the kinetic energy of the basic flow, see figure \ref{Fig_TDEIm_eps02} (b).
This translates to a dimensional flow amplitude $\epsilon\,\Omega_s R_*$ regardless of the strong stratification.
The total temperature field displayed in figure \ref{Fig_Ttot} (c) seems however hardly disturbed by the instability, implying that the motions are mostly confined to spherical shells with almost no radial component.
This is confirmed by the ratio $F_{pol}$ of poloidal kinetic energy to the total kinetic energy, shown in figure \ref{Fig_Fpol}.
For $N_0/\Omega_s \leq 1$, $F_{pol}$ mostly lies between $0.3$ and $0.4$.
When $N_0/\Omega_s \geq 1$, first $F_{pol}$ seems to take values between $0.1$ and $0.5$, before dropping below $0.05$ when the stratification is further increased in the range $N_0/\Omega \geq 10$.
These low values of the poloidal kinetic energy show that the flow has consistently a weak radial component when $N_0/\Omega_s \geq 10$.

	\subsection{Kinematic dynamos}
\begin{figure}
	\centering
	\includegraphics[width=0.45\textwidth]{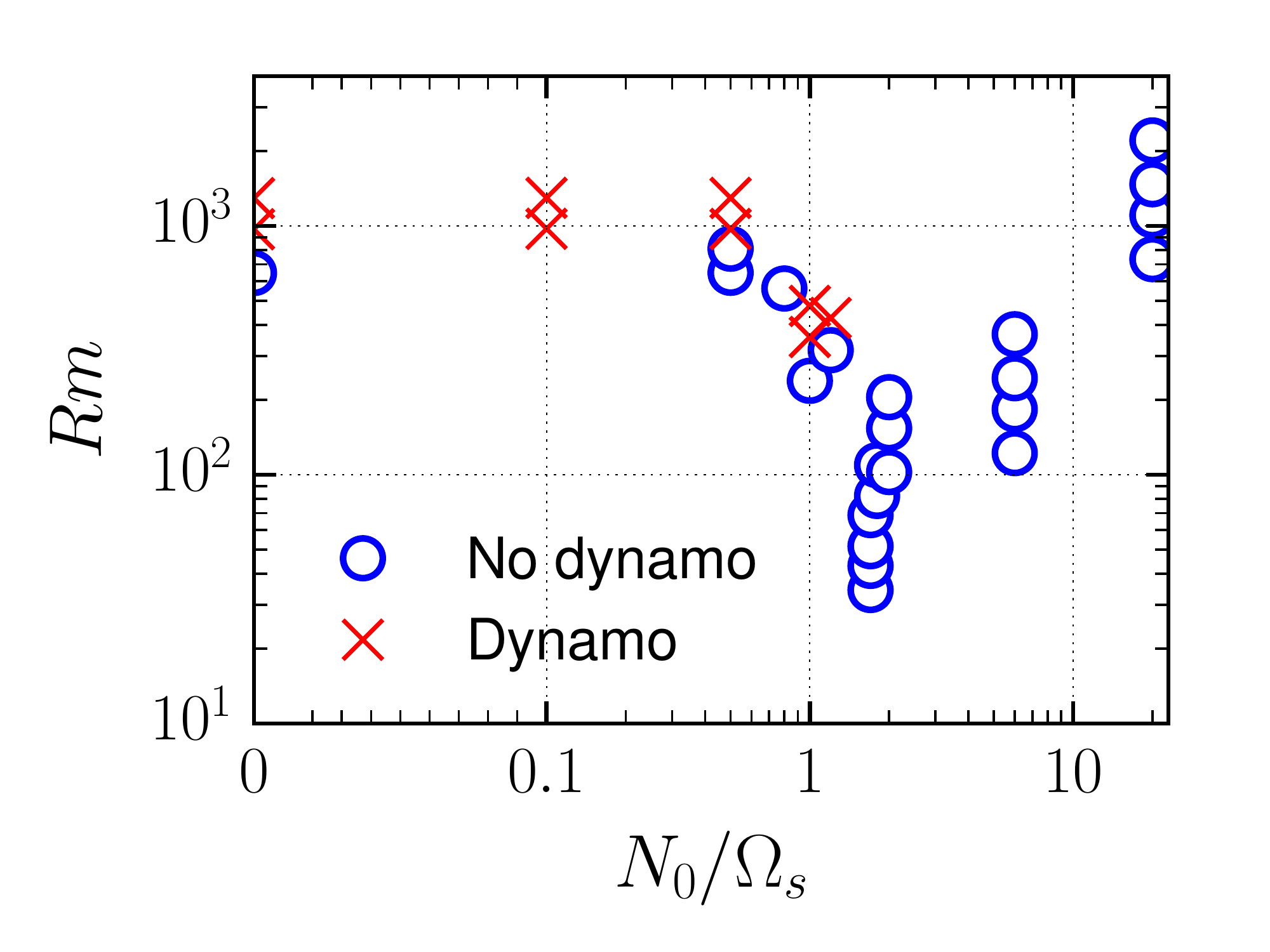}
	\caption{Survey of kinematic dynamos for varying $N_0/\Omega_s$ and $Rm$. Simulations at $Ek=10^{-4}, Pr=1$ and $\epsilon=0.2$}
	\label{Fig_KineDyn_RmN0_eps015}
\end{figure}

We remove the Lorentz force $(\boldsymbol{\nabla} \times \boldsymbol{B}) \times \boldsymbol{B}$ from the momentum equation  (\ref{eq:goveqnU}) to investigate kinematic dynamos. In this problem, we assess the dynamo capability of the nonlinear tidal motions, without a back reaction of the magnetic field on the flow. We introduce the magnetic Reynolds number
\begin{equation}
	Rm = Pm \, Re,
	\label{eq:Rm}
\end{equation}
with $Re$ the Reynolds number previously introduced.
If the structure of the tidal instability flow is suitable for dynamo action, $Rm$ has a finite critical value $Rm_c$ above which the dynamo process starts, characterized by the growth of a magnetic field. Equivalently, the dynamo threshold $Rm_c$ is associated with a critical magnetic Prandtl number $Pm_c$ for a fixed value of $Ek$.

We have considered several values of the magnetic Prandtl number ($1 \lesssim Pm \leq 5$), starting from random magnetic seeds, to determine $Pm_c$. We have checked that the laminar basic flow $\boldsymbol{U}_0$ is not dynamo capable for $Pm \leq 5$, but it does not preclude a laminar dynamo driven by the basic flow at higher $Pm$. 
To detect the onset of kinematic dynamo action, we monitor the time-evolution of the mean magnetic energy density $E(\boldsymbol{B}) = \int_V |\boldsymbol{B}|^2/2 \, \mathrm{d}V$ and deduce the growth rate $\sigma_b$ by fitting with an exponential function.
The kinematic dynamos we obtained are summarized in figure \ref{Fig_KineDyn_RmN0_eps015}. Typical growth rates are $\sigma_b = \mathcal{O}(10^{-3})$.

Nonlinear motions are always dynamo capable when $0 \leq N_0/\Omega_s\lesssim 1$, at least for $Pm \geq 1.5$ at $Ek=10^{-4}$. This yields a typical dynamo threshold $Rm_c \simeq 3000$, a plausible value for dynamo action. This value is higher than the one obtained for precession-driven \citep{tilgner2005precession,goepfert2016dynamos} and tidally driven \citep{cebron2014tidally} dynamos in neutral fluids.

In the range $1 \lesssim  N_0/\Omega_s\leq1.3$, several dynamos are obtained with a smaller $Rm_c \simeq 1000$.
In the range $1.3 \leq N_0/\Omega_s < 10$, no dynamo is obtained for the considered $Pm \leq 2$.
This is because the saturated amplitude of the flow is weak ($Re \simeq 100$), as a result of a higher $\epsilon_c$ there, leading to a much lower supercriticality (see appendix \ref{sec:append_weakening}).
Studying this region would require a more systematic parameter survey, and in particular lowering the diffusivities. This would require more computational power than we currently have at our disposal and this is left for a future study.
For stronger stratifications ($N_0/\Omega_s \geq 10$) we found no dynamo, even for the most extreme case with $Rm \simeq 8000$. This suggests that the nonlinear tidal flows in this range are not dynamo capable as a results of their spatial structure, even if the Reynolds number can be larger ($Re \leq 2000$).
Indeed, the toroidal velocity theorem states that an incompressible flow without radial component (i.e. purely toroidal) cannot sustain a magnetic field \citep{bullard1954homogeneous}. This theorem is not invalidating when small non-radial motions are considered \citep{kaiser2017robustness}.
When $N_0 \gg \Omega_s$, although being of considerable amplitude, the tidally driven flow seems constrained by the stratification and leads to weak radial motions, see figure \ref{Fig_Ttot} (b). This is a plausible explanation of the lack of dynamos for reasonable values of $Rm$ at $N_0/\Omega_s > 10$.

	\subsection{Self-consistent dynamos}
\begin{figure}
	\centering
	\subfloat[]{
		\includegraphics[width=0.48\textwidth]{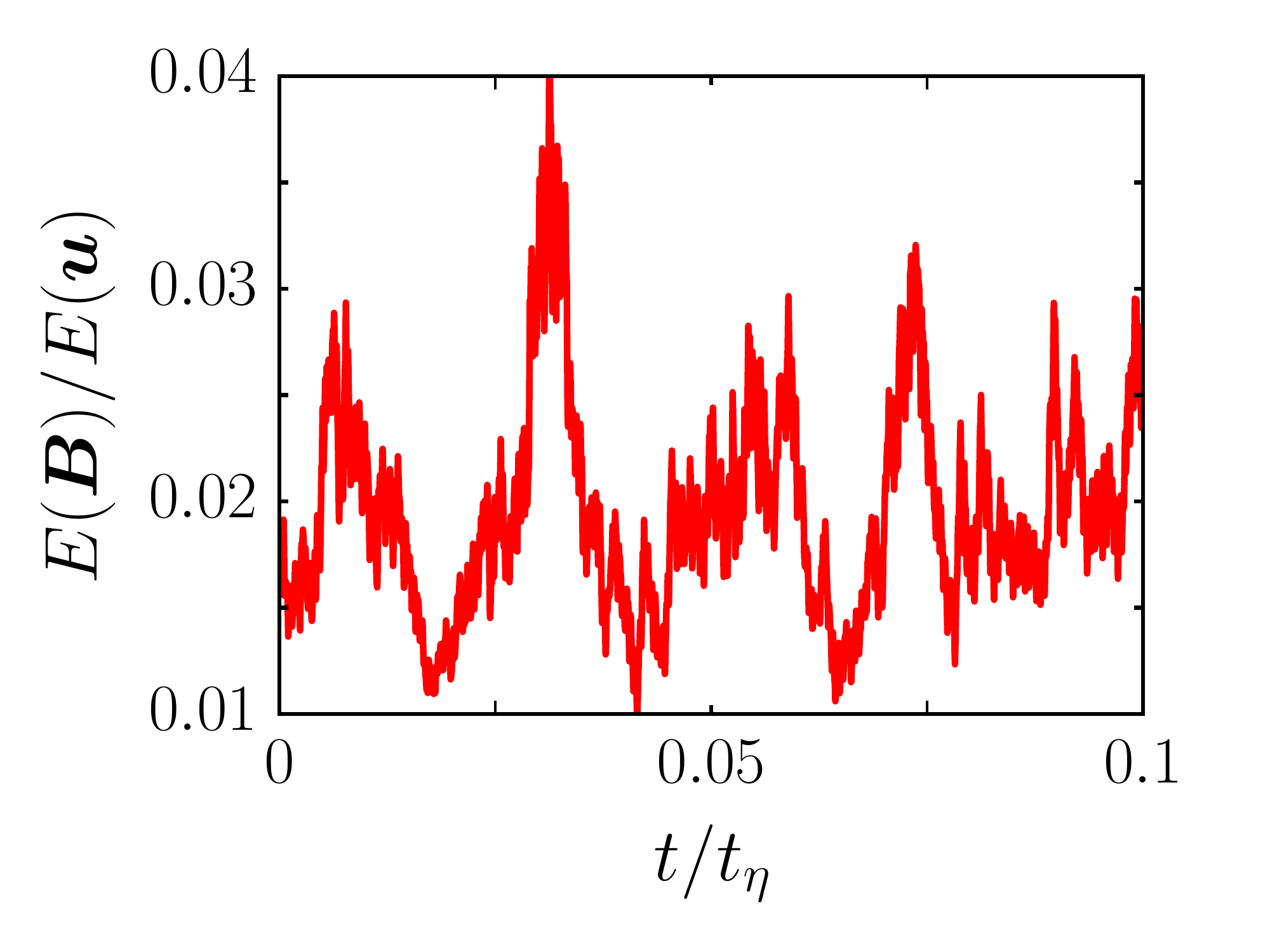}
		}
		\\
	\subfloat[]{
		\includegraphics[width=0.48\textwidth]{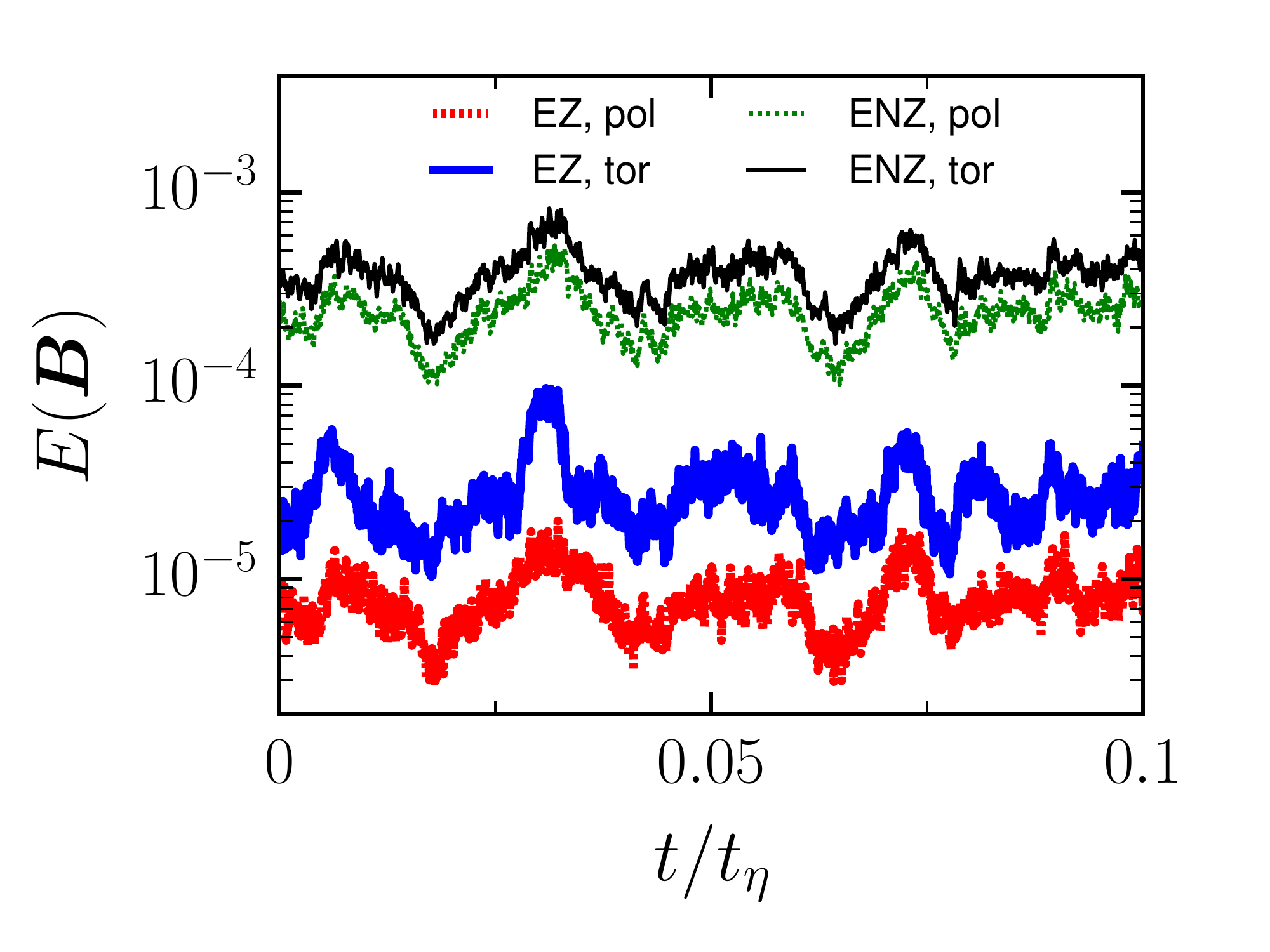}
		}
	\caption{Self-consistent magnetic field in the saturation regime. Simulation at $Ek=10^{-4}, Pr=1, Pm=2, \epsilon=0.2$. Only a small representative fraction of the dimensionless diffusive time $t_{\eta} = Pm/Ek$ is shown (a) Ratio $E(\boldsymbol{B})/E(\boldsymbol{u})$. (b) Poloidal zonal ($m=0$) energy (EZ, pol), toroidal zonal ($m=0$) energy (EZ, tor), poloidal non-zonal ($m>0$) (ENZ, pol) energy and toroidal non-zonal ($m>0$) energy (ENZ, tor) of the magnetic field.}
	\label{Fig_EB}
\end{figure}

Now we take the Lorentz force into account in the momentum equation (\ref{eq:goveqnU}) to compute self-consistent dynamos.
We integrate the governing equations (\ref{eq:goveqn}) over one dimensionless magnetic diffusive time $t_{\eta} = Pm/Ek$ to get reliable dynamo results.
We use the saturated tidal flow as initial conditions for the velocity field.
All the kinematic dynamos obtained for $N_0/\Omega_s \lesssim 1$ give self-consistent dynamos.
As in the hydrodynamic case, the simulations are qualitatively and quantitatively similar in the whole range $N_0/\Omega_s \lesssim 1$. We only provide a detailed analysis of the illustrative simulation performed at $N_0/\Omega_s = 0.5$, $\epsilon=0.2$ and $Pm=2$ (with $1.25<Pm_c<1.5$).

The magnetic energy, initially weak, is amplified and reaches values representing a small fraction of the kinetic energy of the flow driven by the tidal instability in figure \ref{Fig_EB} (a).
This fraction is about $0.01 - 0.02$.
Hence, the magnetic field does not reach a state of equipartition and the kinetic energy is therefore only slightly affected by the dynamo action.
Note that these values are smaller than those obtained by \citet{barker2013nonb} in local simulations without buoyancy effects.
However, with larger $Rm$, larger amplitude of the magnetic energy could be reached.
The time evolution of the magnetic field seems to follow the time evolution of the velocity field (see figure \ref{Fig_EB}).
Magnetic energy has rapid oscillations, at frequency of the order of the spin rate, which are superimposed on longer period oscillations of small amplitudes.
In figure \ref{Fig_EB} (b), we observe that the zonal energy (i.e. axisymmetric $m=0$ energy) is one order of magnitude smaller than the non-zonal energy (i.e. non-axisymmetric $m>0$ energy). The magnetic field is also predominantly toroidal, as expected from stability considerations in non-barotropic stars \citep{akgun2013stability}.

\begin{figure}
	\centering
	\subfloat[]{
		\includegraphics[width=0.45\textwidth]{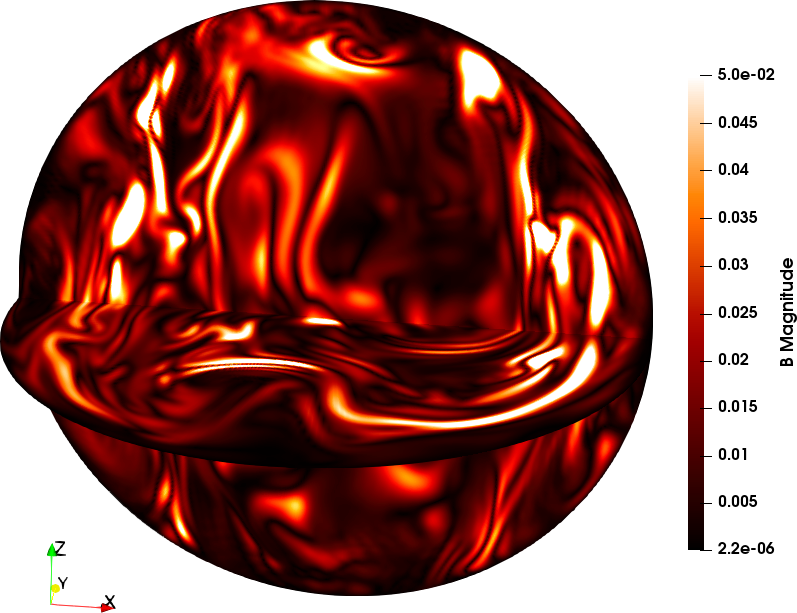}
		}
		\\
	\subfloat[]{
		\includegraphics[width=0.48\textwidth]{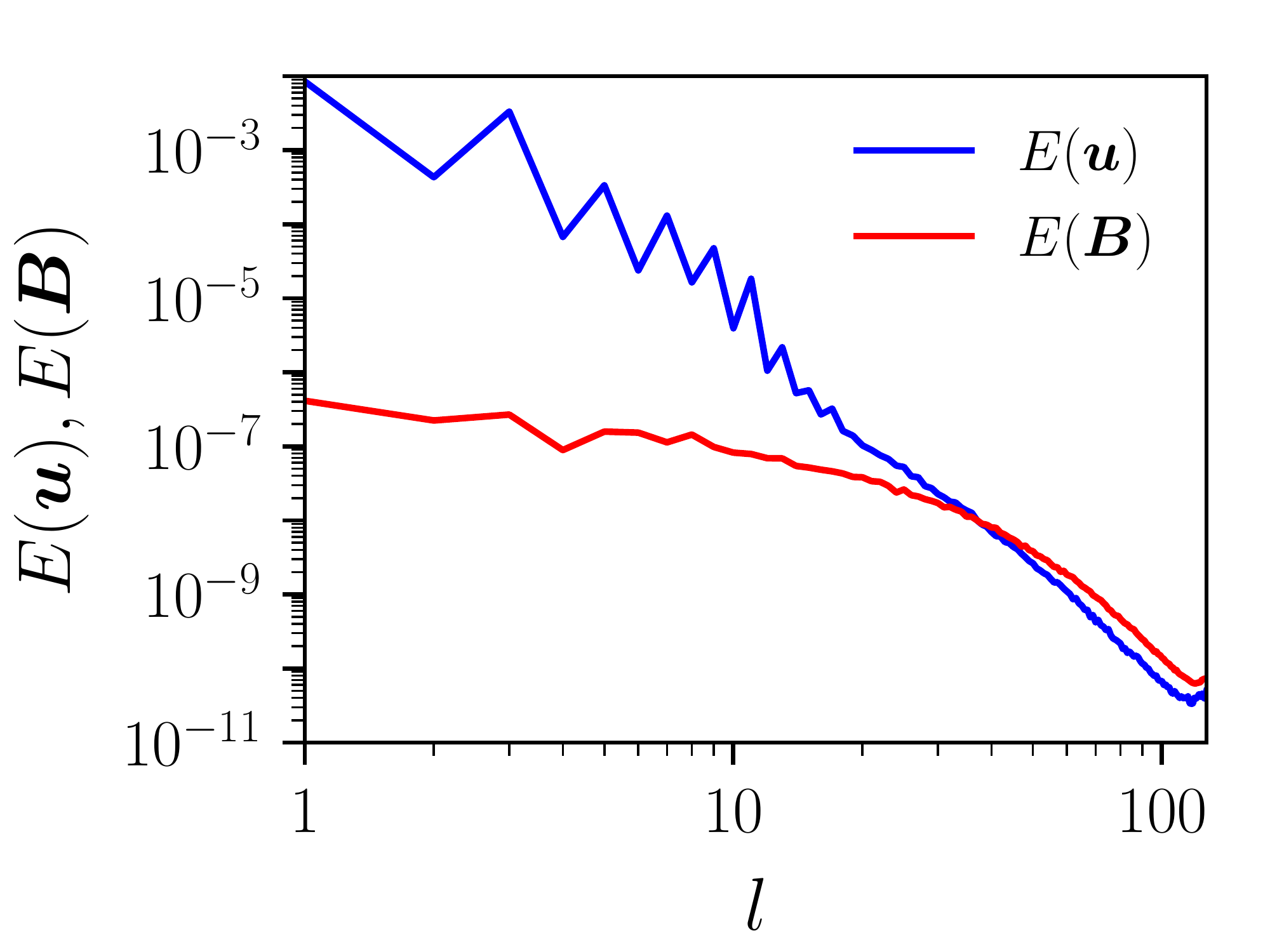}
		}
	\caption{(a) Three-dimensional snapshot of the magnetic field magnitude $|\boldsymbol{B}|$ at a given time. A movie showing the time evolution of the field is given in the supplementary materials. The rotation axis is along $z$.  (b) Time and radius averaged spectra of the magnetic energy as function of the spherical harmonic degree $l$. Simulation at $Ek=10^{-4}, Pr=1, Pm=2$ and $\epsilon=0.2$.}
	\label{Fig_B_eps02_N05}
\end{figure}

Because of the complex time evolution, straightforward visualisations of the instantaneous field are not illuminating. We show in figure \ref{Fig_B_eps02_N05} (a) an instantaneous snapshot of the magnitude of the magnetic field. The field is of rather small scale.
We observe similarities with the temperature field shown in figure \ref{Fig_Ttot} (a). 
A description of the field morphology is provided by the time averaged spectrum of the magnetic field in figure \ref{Fig_B_eps02_N05} (b).
The magnetic spectrum is dominated by components of spherical harmonic degrees $l\leq10$.
It is maximum for the dipolar component ($l=1$) and then slowly decays with a power-law  $E(\boldsymbol{B}) \propto l^{-0.04}$. 
The time-averaged spectrum, as well as the instantaneous ones, are well-resolved, proving that tidal motions are able to drive a dipole-dominated dynamo in a stably stratified layer.

\begin{figure}
	\centering
	\subfloat[Radial magnetic field]{
		\includegraphics[width=0.48\textwidth]{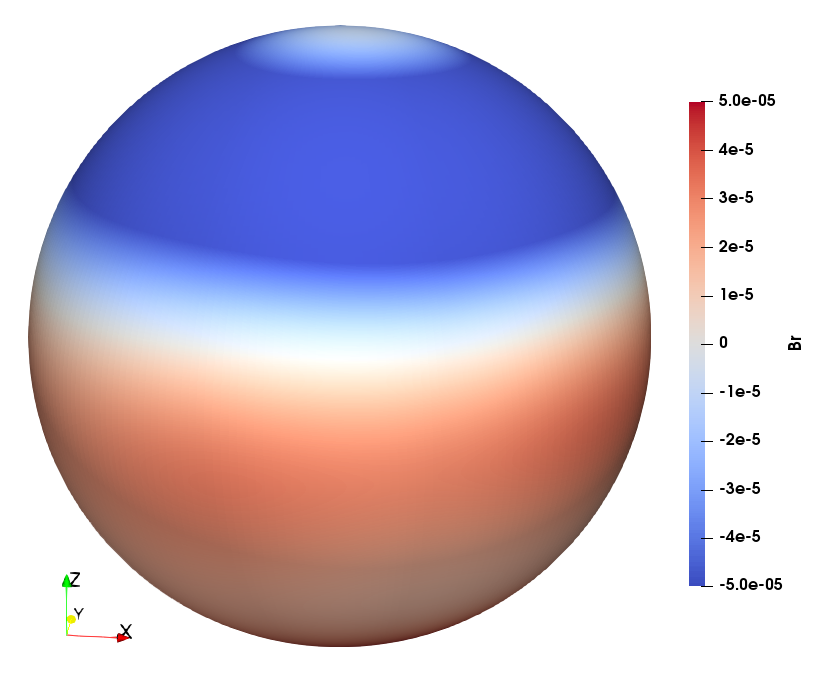}
		}
	\\
	\subfloat[Velocity magnitude $|\boldsymbol{u}|$]{
		\includegraphics[width=0.48\textwidth]{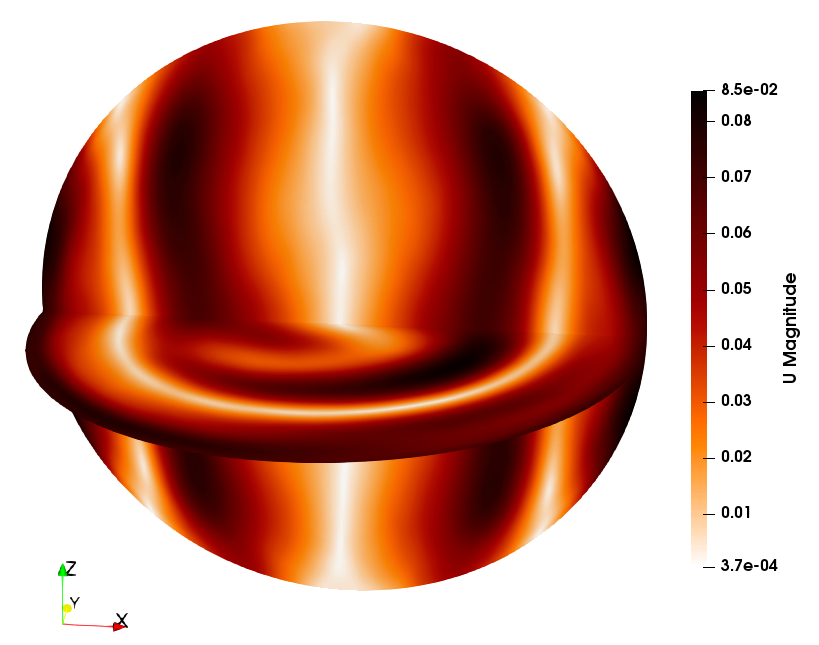}
		}
	\caption{(a) Time-averaged radial magnetic field at the stellar surface and (b) time-averaged velocity magnitude in the equatorial plane and in a meridional plane. Simulations at $Ek=10^{-4}, Pr=1, Pm=2$ and $\epsilon=0.2$. Time-averaged fields computed from $t/t_\eta=0$ to $t/t_\eta=0.1$ in figure \ref{Fig_EB} (b). In both figures the spin axis is the vertical $z$ axis.}
	\label{fig:surfaceB}
\end{figure}

We show in figure \ref{fig:surfaceB} the time-averaged magnetic field truncated at spherical harmonics degree  $l=5$, because higher degrees are not observed \citep[e.g.][]{donati2009magnetic,fares2017moves}.
This time-averaged field is mostly dipolar ($l=1$) and axisymmetric ($m=0$). Non-axisymmetric components are averaged out because of the rapid spin.
The time-averaged flow has a columnar structure aligned with the spin axis, as shown in figure \ref{fig:surfaceB} (b).
These spin-aligned structures are the global counterpart of
the strong vortices almost invariant along the rotation axis and filling the periodic boxes of similar local simulations   \citep{barker2013nona,barker2013nonb}.
These flows are expected in our stress-free computations with no viscous friction at the boundary \citep{livermore2016comparison,le2017inertial}.
The emergence of such spin-aligned large-scale vortices are also observed in rotating thermal convection \citep[e.g.][]{guervilly2014large} and have been shown to be dynamos \citep{guervilly2015generation}.

	\subsection{Tidal mixing}
\begin{figure}
	\centering
	\subfloat[]{
		\includegraphics[width=0.48\textwidth]{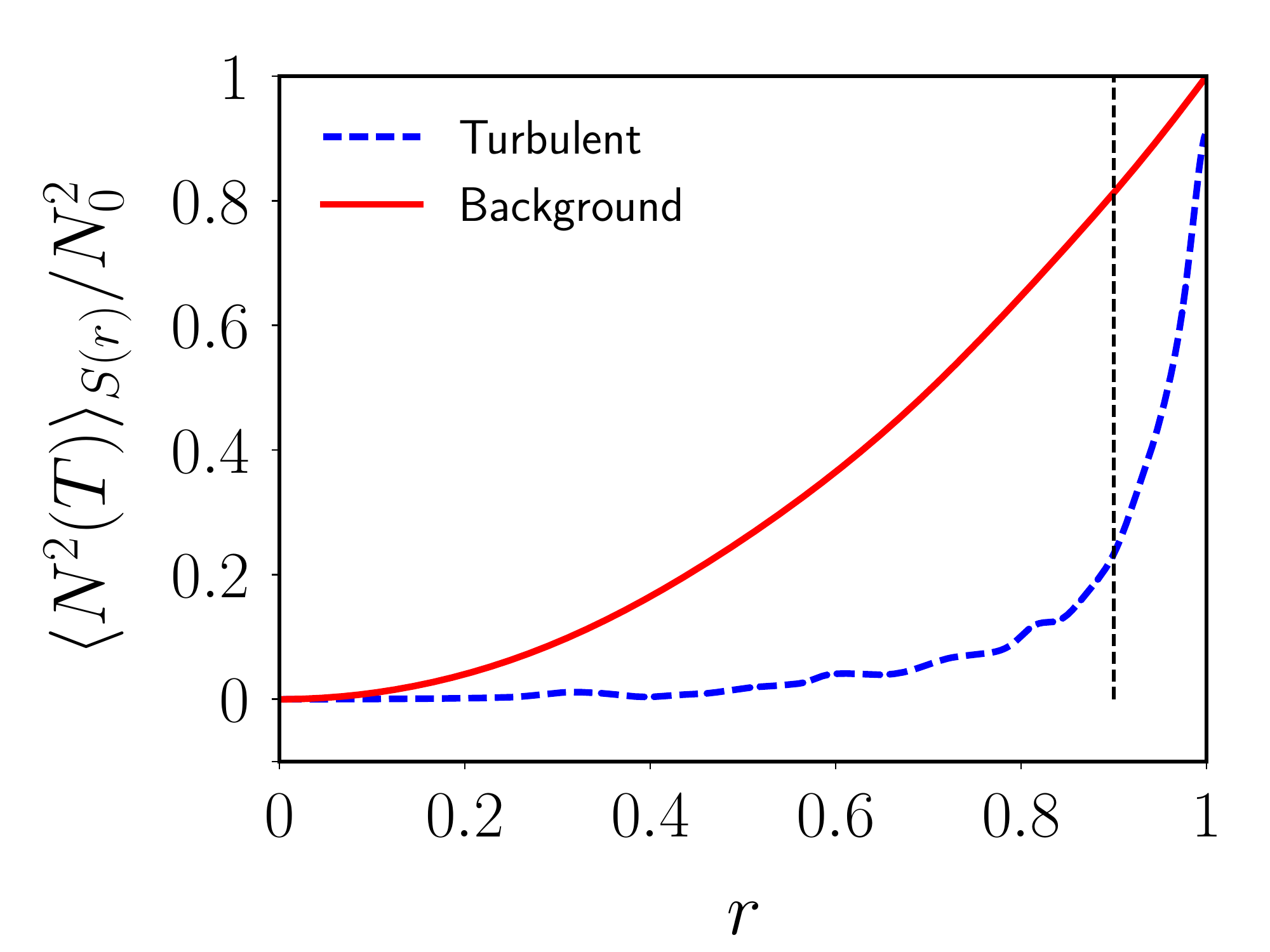}
		}
	\\
	\subfloat[]{
		\includegraphics[width=0.48\textwidth]{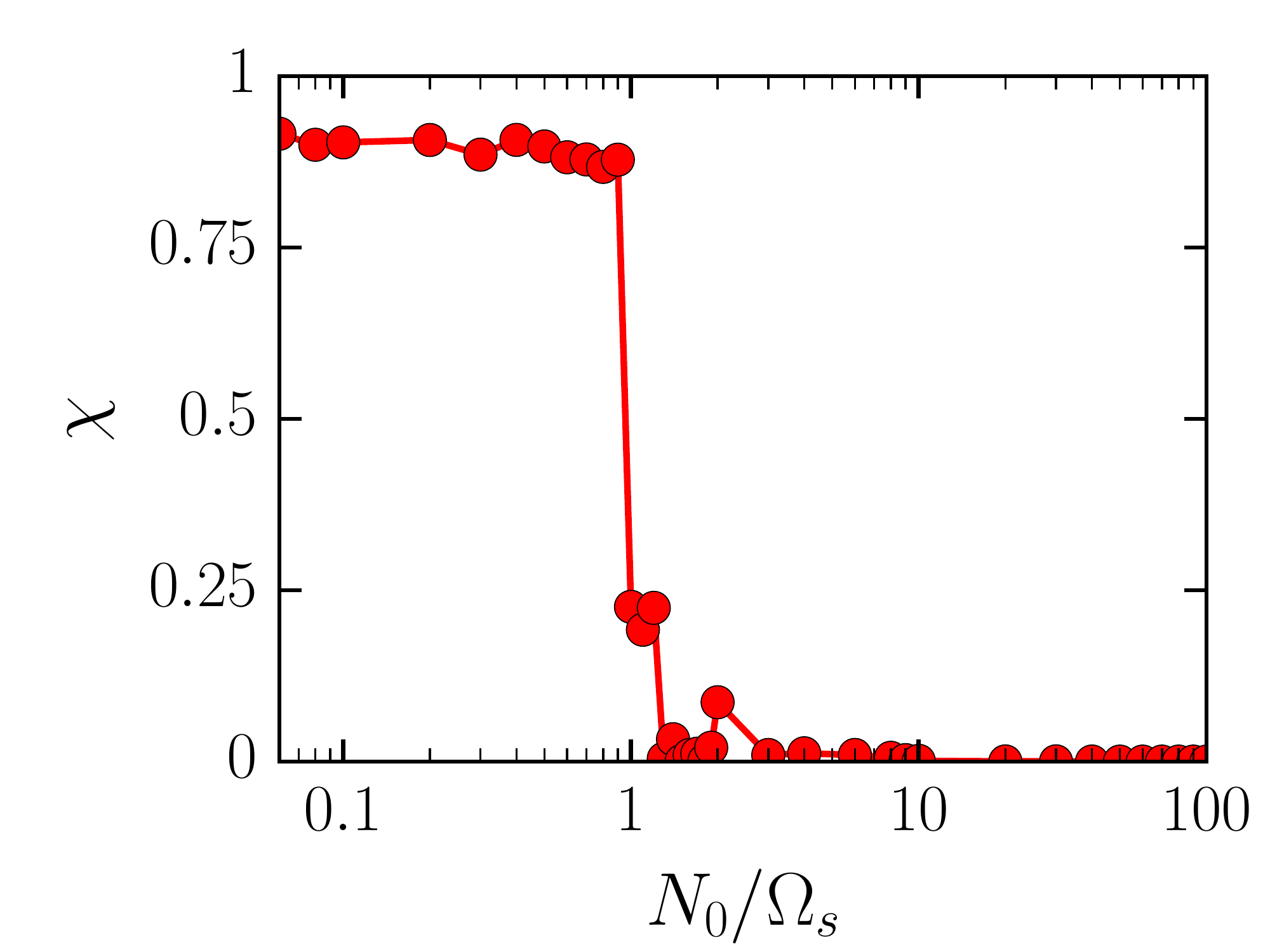}
		}
	\caption{(a) Time average of the surface average ($l=0,m=0$) of the local Brunt-V\"ais\"al\"a frequency $\langle N^2 (T) \rangle_{S(r)}$ as function of radius $r$. The vertical dashed line represents the beginning of the thermal boundary layer. (b) Efficiency of mixing $\chi$ for varying $N_0/\Omega_s$. We fix $r_\text{tbl} = 0.9$ in formula (\ref{eq:mixing}). Simulations at $N_0/\Omega_s = 0.5, Ek=10^{-4}, Pr=1$ and $\epsilon=0.2$}
	\label{Fig_Mixing}
\end{figure}

We have shown that the tidal instability is dynamo capable in our simulations when $N_0/\Omega_s \lesssim 1$ with a dynamo threshold $Rm_c \simeq 3000$. 
For stronger stratifications ($N_0/\Omega_s \geq 10$), we did not find dynamo action up to $Rm \simeq 8000$ in the simulations. 
Indeed, dynamo action requires not only large $Rm$, but also adequate, sufficiently complex, flow structure \citep{kaiser2017robustness}. Here, we suspect the radial mixing induced by the tidal forcing to be important. Therefore, we now quantify the mixing induced by nonlinear tidal motions.

As shown in figure \ref{Fig_EB} (a), the magnetic energy is much smaller than the kinetic energy. Hence, the Lorentz force has little effect on the flow dynamics. To quantify how the background temperature $T_0$ is mixed by the tidal instability, we compute the time and spherical average of the local Brunt-V\"ais\"al\"a frequency  $\langle N^2 (T) \rangle \,_{S(r)}$, where  $\langle . \rangle \,_{S(r)}$ is the average over the spherical surface $S(r)$ at radius $r$ (i.e. $l=0$ in spectral space).
It is illustrated in figure \ref{Fig_Mixing} (a) for the nonlinear saturated regime of the simulation at $N_0/\Omega_s = 0.5$ and $\epsilon=0.2$ (representative of the stratification $N_0/\Omega_s \leq 1$).
The dashed line represents the background state. In the nonlinear state (dashed line), the stratification is well-mixed ($N^2 (T) \simeq 0$) as suggested by figure \ref{Fig_Ttot} (a), except near the outer boundary where a thermal boundary layer appears. 
This thermal boundary layer has a typical thickness of about $0.1$ in our simulations.

To estimate the efficiency of the mixing, we compute a coefficient of mixing $\chi$ defined as follows
\begin{equation}
	\chi = \left | 1 - {\int_0^{r_\text{tbl}} \langle N^2 (T) \rangle_{S(r)} \mathrm{d} r} \left (\int_0^{r_\text{tbl}} \langle N^2 (T_0) \rangle_{S(r)} \mathrm{d} r \right )^{-1} \right |,
	\label{eq:mixing}
\end{equation}
with $r_\text{tbl}=0.9$ the bottom radius of the thermal boundary layer. If $\chi=1$ then the stratification is entirely mixed (below the thermal boundary layer), while if $\chi=0$ there is no mixing. Figure \ref{Fig_Mixing} (b) displays the evolution of $\chi$ with $N_0/\Omega_s$. We find that the stratification is  almost entirely mixed by the tidal instability (below the thermal boundary layer) when $N_0/\Omega_s \lesssim 1$. When $1 \lesssim N_0/\Omega_s \leq 2$, the mixing efficiency is strongly reduced. 
Then, we find that there is no mixing associated with the still vigorous tidal motions when $N_0/\Omega_s \geq 2$. 
We explain the observed dichotomy below and above $N_0/\Omega_s = 1$ based on the following simple arguments. A parametric resonance involving inertial modes is responsible for the tidal instability, which is almost insensitive to the stratification when $N_0/\Omega_s\lesssim 1$. 

When $1 \lesssim N_0/\Omega_s \leq 2$, Coriolis and buoyancy forces are of the same order and thus a parametric instability involving inertia-gravity modes is responsible for the tidal instability. 
However, as shown in appendix \ref{sec:append_weakening}, the collapse in the kinetic energy in figure \ref{Fig_TDEIm_eps02} (b) when $1 \lesssim  N_0/\Omega_s \leq 2$, responsible for the strong reduction of the mixing in figure \ref{Fig_Mixing} (b), is due to a higher $\epsilon_c$ and to a lower supercriticality there.
It is not expected to occur in stellar interiors in the asymptotic limit $Ek, Ek/Pr \to 0$ for these values of $N_0/\Omega_s$. Thus, for smaller $Ek$, radial mixing is also expected in nonlinear regimes.

Finally for stronger stratifications ($N_0/\Omega_s \geq 2$), the tidal instability generates motions mainly along spherical shells, as indirectly observed in the advection of the scalar temperature in figure \ref{Fig_Ttot} (c).
The tidal instability is linearly triggered near the locus of maximum ellipticity ($r=0.5$) and generates there nonlinear radial motions of short wavelengths (not shown).
This is because the ellipticity is not homogeneous in our model (see figure \ref{fig:Ellipticity_Meridional}), 
but in a ellipsoidal body (like a tidally deformed star) we expect it to appear everywhere.
Nonlinear motions are mostly toroidal motions of spherical coefficients ($l=1,m=1$). These motions seem similar to "r-modes"-like motions, which are the least damped motions with stress-free boundary conditions \citep{rieutord2001ekman}.
The strong stratification inhibits radial flows and toroidal flows are favoured instead, unable to lead to efficient radial mixing.

\section{Astrophysical applications}
\label{sec:astrophysics}
\begin{figure}
	\centering
	\includegraphics[width=0.45\textwidth]{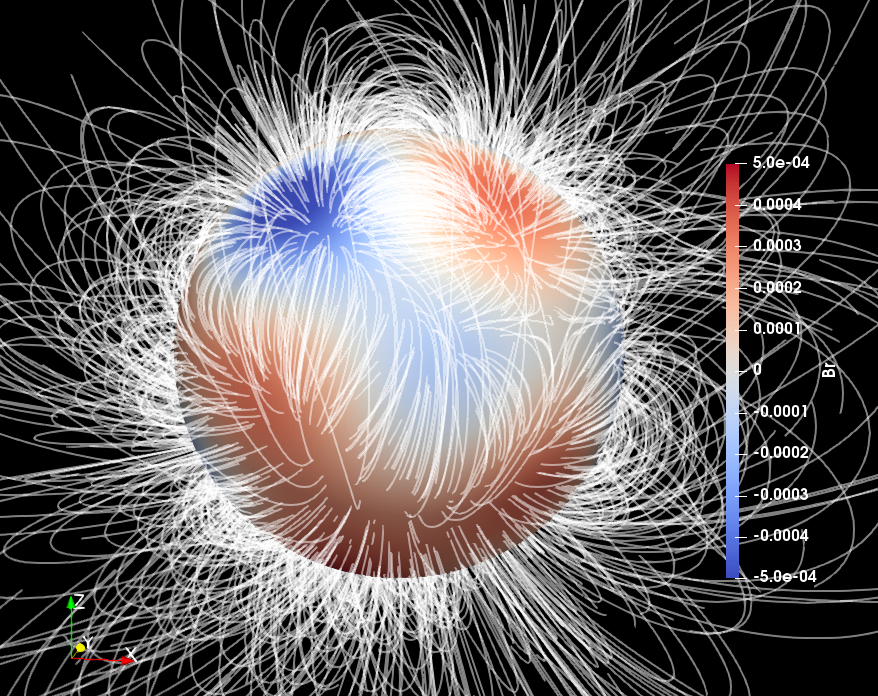}
	\caption{Potential field extrapolation of the instantaneous surface magnetic field (up to dimensionless radius $r=2$). Simulations at $Ek=10^{-4}, Pr=1, Pm=2$ and $\epsilon=0.2$.}
	\label{fig:surfaceBdipol}
\end{figure}

To investigate the astrophysical importance of the tidal instability for stellar magnetism, we have to extrapolate our numerical results towards the parameter space of stellar interiors. 
We expect our numerical simulations to capture the dominant global scales of tidally driven nonlinear motions. 
Indeed, there is a broad agreement with the observed magnetic pattern at the surface of many magnetic stars, showing a dominant dipolar field with possible smaller scales \citep{donati2009magnetic}. The instantaneous magnetic field and the potential field extrapolation (external field) of a model are shown in figure \ref{fig:surfaceBdipol}, truncating the magnetic spectrum at $l\leq5$. Higher harmonics are not observed in astronomical data. The external potential field is still dominated by the dipolar component.
Without scaling laws, we cannot extrapolate towards the parameter space of stellar interiors.
Unfortunately, all available scaling laws have been developed for convective dynamos only \citep[e.g.][]{christensen2009energy,yadav2013scaling,yadav2013consistent,augustson2016dynamo} and cannot be safely applied to other forcings.
Obtaining scaling laws would require to simulate lower viscosities, which are currently out of reach.

\subsubsection*{Tidal instability in stellar interiors}

We carry out the extrapolation as follows. We consider a star of mass $M_*$, mean radius $R_*$ and equatorial ellipticity $\epsilon_*$. The radiative zone is modelled as a stably stratified zone in the Boussinesq approximation. A tidal basic flow (equilibrium tide), induced by the disturbing tidal potential of an orbiting companion of mass $m$, is established within the radiative envelope. We consider only non-synchronised systems, where the spin angular velocity of the star $\Omega_s=2\pi/P_s$ (with $P_s$ the spin period) is not equal to the mean orbital rotation rate of the companion $\Omega_\text{orb}=2\pi/P_\text{orb}$ (with $P_\text{orb}$ the orbital period). For simplicity we assume that the companion is moving on a circular orbit in the equatorial plane of the host star. The ellipticity $\epsilon_*$ is estimated from the static bulge theory \citep[e.g.][]{cebron2012elliptical,vidal2017inviscid} 
\begin{equation}
	\epsilon_* = \frac{3}{2} \frac{m}{M_*} \left ( \frac{R_*}{D} \right )^3,
	\label{eq:estimate_beta1}
\end{equation}
with $D$ the typical distance between the star and its orbital companion. An estimation of $D$ can be obtained with Kepler's third law, yielding \citep{barker2013nonb}
\begin{equation}
	\epsilon_* = \frac{3}{2} \frac{m}{m + M_*} \left ( \frac{\Omega_\text{orb}}{\Omega_\text{dyn}} \right )^2,
	\label{eq:estimate_beta2}
\end{equation}
with the dynamical frequency $\Omega_\text{dyn} = \sqrt{GM_* / R_*^3}$ and $G$ the gravitational constant. 

The fastest growing mode of the tidal instability (in the asymptotic limit $Ek,Ek/Pr \to 0$) has the dimensional growth rate \citep[e.g.][]{kerswell2002elliptical}
\begin{equation}
	\frac{\sigma}{|\Omega_s-\Omega_\text{orb}|} = \frac{(2 \widetilde{\Omega}+3)^2}{16(1+\widetilde{\Omega})^2} \epsilon_*,
	\label{eq:sigma_TDEI}
\end{equation}
with $\widetilde{\Omega} = \Omega_\text{orb}/(\Omega_s - \Omega_\text{orb})$ the background rotation. Using astronomical quantities, formula (\ref{eq:sigma_TDEI}) is rewritten as
\begin{equation}
	\sigma = \frac{3}{2} \left | 1-\frac{\Omega_\text{orb}}{\Omega_s} \right | \frac{(2 \widetilde{\Omega}+3)^2}{16(1+\widetilde{\Omega})^2} \frac{m}{D^3} \frac{R_*^3  \Omega_s}{M_*} \leq 3 \frac{m}{D^3} \frac{R_*^3  \Omega_s}{M_*}.
	\label{eq:sigma_TDEI2}
\end{equation}
The growth rate (\ref{eq:sigma_TDEI2}) is insensitive to the amplitude of the stratification $N_0/\Omega_s$, as globally observed in our simulations (except for $1 \lesssim N_0/\Omega_s \leq 2$, see appendix \ref{sec:append_weakening}).
The tidal instability is triggered for circular orbital configurations belonging to the allowable range $-1 \leq \Omega_\text{orb}/\Omega_s \leq 3$ \citep[e.g.][]{le2010tidal}. However, the tidal instability can be excited well outside this range for eccentric Kepler orbits \citep{vidal2017inviscid}.

Based on our global simulations of the tidal instability, buoyancy effects do not influence amplitudes of tidal nonlinear motions when $N_0/\Omega_s  \leq 1$.
For $N_0/\Omega_s \geq 10$, the tidal instability stays vigorous but the flow is constrained by the strong stratification resulting in weak radial motion, see \S\ref{sec:hydro} and figure \ref{Fig_Mixing}.
When $1 < N_0/\Omega_s < 10$, the lower amplitudes observed are due to a larger critical ellipticity, see appendix \ref{sec:append_weakening}.
Therefore, as shown in figure \ref{Fig_TDEIm_eps02} (b), the tidal instability generates nonlinear flows with a typical velocity magnitude \citep{barker2013nona,barker2013nonb,grannan2016tidally}
\begin{equation}
u \sim \epsilon_* |\Omega_s - \Omega_\text{orb}| R_*.
\label{eq:ampl_U}
\end{equation}

\subsubsection*{Prediction for the magnetic field strength}

Dynamo action requires a large magnetic Reynolds number, i.e. $Rm > Rm_c$.
This translates into a constraint on the magnetic diffusivity
$\eta < u R_* / Rm_c$.
Using the estimate (\ref{eq:ampl_U}) for $u$, we have
\begin{equation}
	\eta < \epsilon_* |\Omega_s - \Omega_\text{orb}| R_*^2 / Rm_c.
	\label{eq:eta_rm}
\end{equation}
We assume a weak dependence of the dynamo threshold $Rm_c$ on $Pm_c$ when the diffusivities are decreased towards stellar values (i.e. $Ek\to0, Pm \ll 1$). Such a behaviour has been reported for several (helical and non-helical) forcing geometries \citep{brandenburg2001inverse,ponty2004simulation,ponty2005numerical,mininni2005dynamo,mininni2007inverse,ponty2007dynamo,brandenburg2009large,seshasayanan2017transition} and seems rather generic.
For $\Omega_s \simeq 1 \, \textrm{d}^{-1}$, $R_* \simeq 2R_\odot$, and $Rm_c=3000$, we obtain
$\eta \lesssim 500 \, \mathrm{m}^2.\mathrm{s}^{-1}$ for $\epsilon_* = 10^{-8}$ and $\eta \lesssim 5\times 10^5 \, \mathrm{m}^2.\mathrm{s}^{-1}$ for $\epsilon_* = 10^{-5}$. The latter values are acceptable values for stellar interiors,. Isuggests that stellar interiors may host dynamo capable flows. 

We relate the dipolar field strength at the stellar surface $B_0$ to the amplitude of the flow (\ref{eq:ampl_U}) using the dimensionless parameter $\delta$ as
\begin{equation}
	B_0 = \delta \: \epsilon_* \sqrt{\rho_* \mu_0} \, |\Omega_s - \Omega_\text{orb}| R_*, \label{eq:scalingB}
\end{equation}
In our simulations, the dipole amplitude at the surface $B_0$ is small compared to the typical magnetic field strength $B_\text{rms}$ within the fluid (see figure \ref{Fig_B_eps02_N05}), leading to $B_0^2 = f_1 E(\boldsymbol{B})$, or $B_0 = \sqrt{f_1} B_\text{rms}$, with $f_1 \simeq 10^{-4}$.
The ratio of the magnetic energy to the kinetic energy is found to be $E(\boldsymbol{B})/E(\boldsymbol{u}) = f_2 = 0.01$ (see figure \ref{Fig_EB}) in our simulations. By contrast, \cite{barker2013nonb} obtained  $f_2 \approx 0.1-0.3$ in their magnetohydrodynamic simulations of the tidal instability within a periodic box. Actually, this ratio largely depends on super-criticality with respect to the dynamo onset.
Equipartition cannot be excluded here in stellar interiors, hence we consider the range $f_2 \in [10^{-2}, 1]$.
This results into $\delta = \sqrt{f_1 f_2} \in [10^{-3},10^{-2}]$. 
Making use of formula (\ref{eq:estimate_beta1}), the scaling law (\ref{eq:scalingB}) can be written using astronomical quantities as
\begin{equation}
	B_0 = \frac{3}{2} \sqrt{\frac{3 \mu_0}{4 \pi}} \delta \, \frac{R_*^{5/2}}{M_*^{1/2}} \Omega_s \frac{m}{D^3} \left | 1 - \frac{\Omega_\text{orb}}{\Omega_s} \right |, 
	\label{eq:scalingB2}
\end{equation}
with the typical density $\rho_* = M_* / (4/3 \pi R_*^3)$.

\begin{figure*}
	\centering
	\begin{tabular}{cc}
	\subfloat[Growth rate]{
		\includegraphics[width=0.48\textwidth]{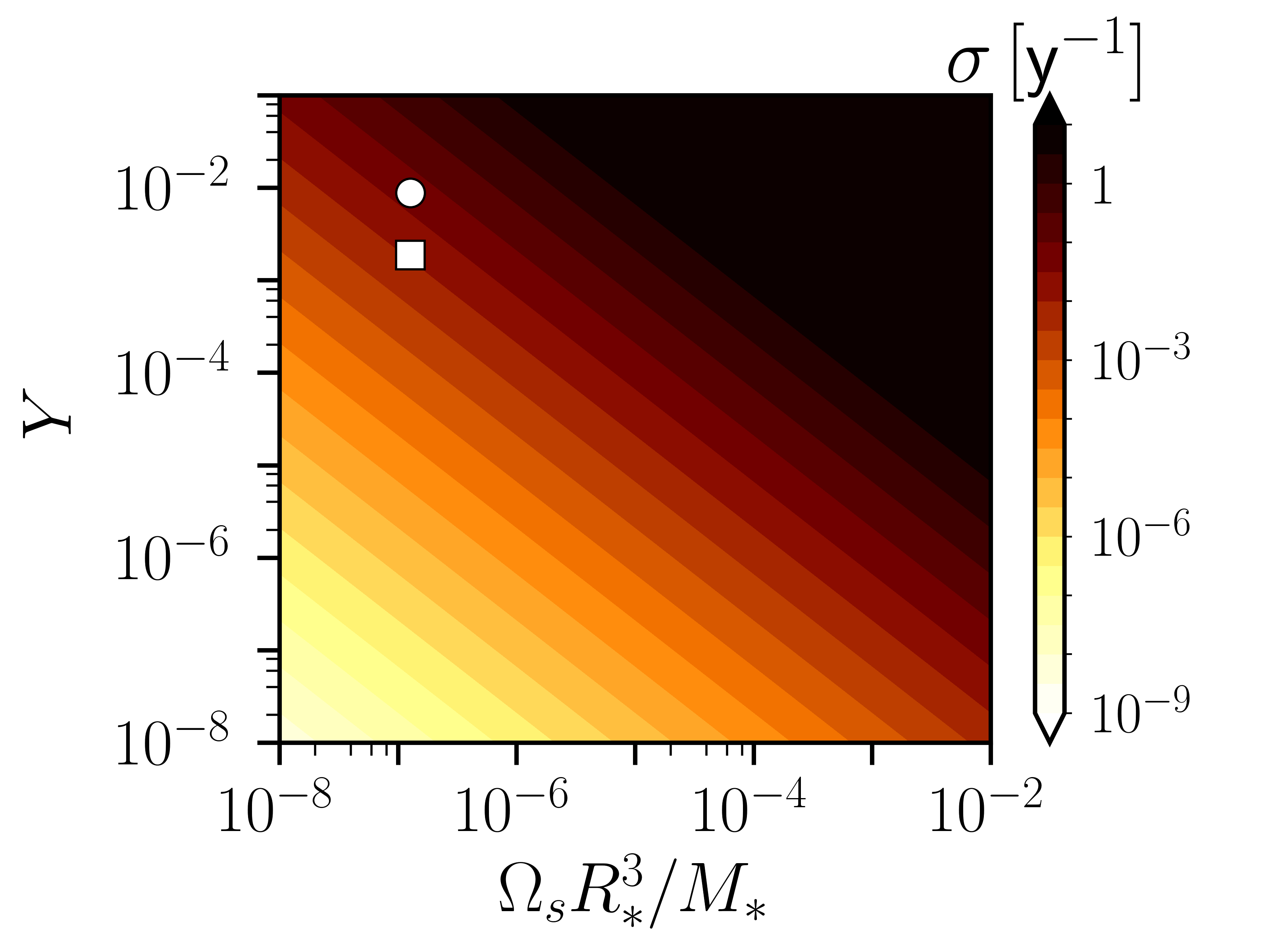}
		}
		&
	\subfloat[Field strength]{
		\includegraphics[width=0.48\textwidth]{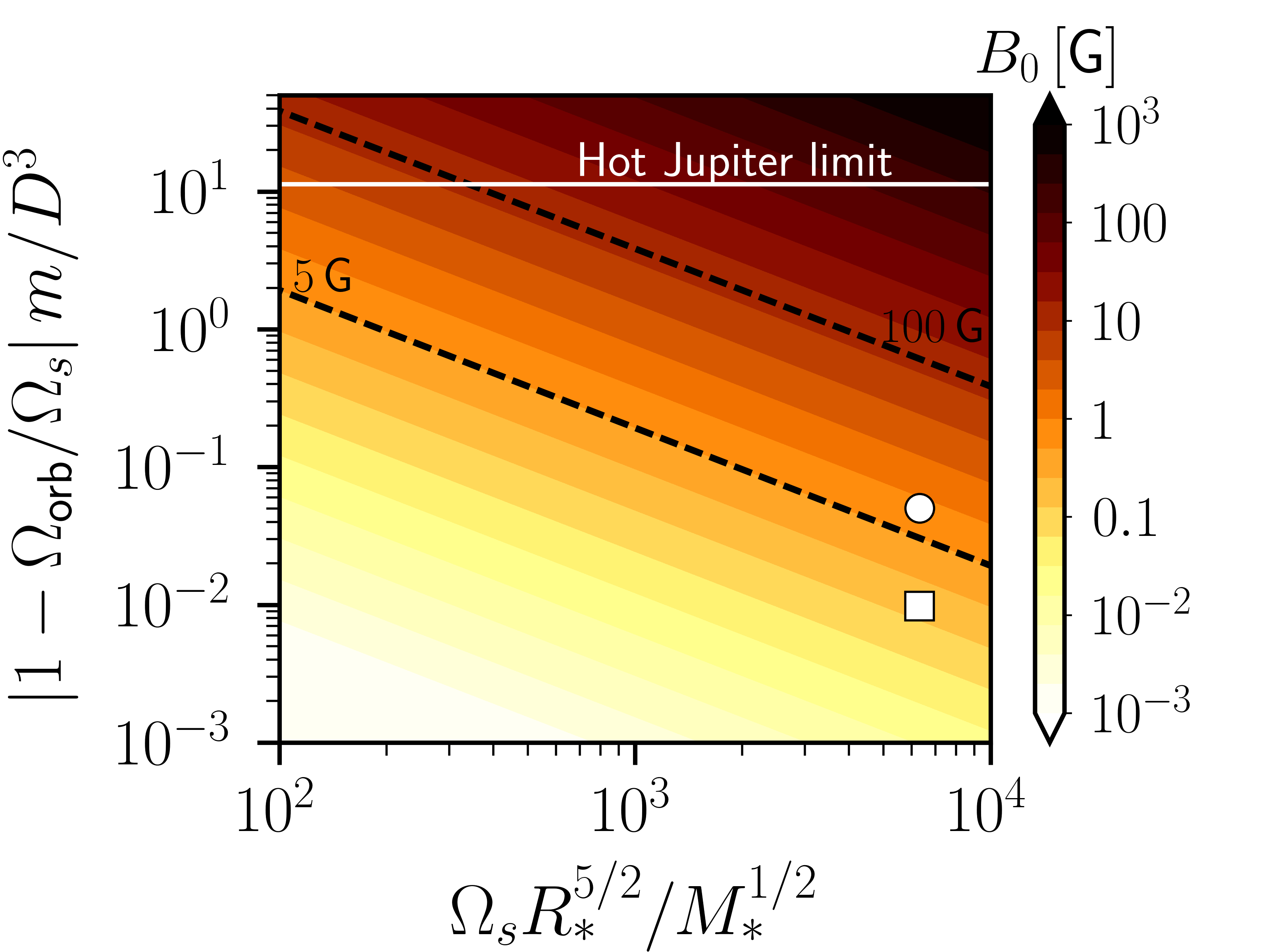}
		}
	\\
	\end{tabular}
	\caption{Predictions of the (a) growth rate (\ref{eq:sigma_TDEI2}) and (b) surface field strength (\ref{eq:scalingB}) with $\delta=10^{-3}$ for various stellar configurations. In (a) the vertical axis shows the quantity $Y=  |1-\Omega_\text{orb}/\Omega_s| (2\widetilde{\Omega}+3)^2/(16(1+\widetilde{\Omega})^2) m/D^3$. Horizontal solid white line in (b) shows the upper limit when the companion is a close and massive Hot Jupiter ($D = 0.01 \, \text{au}, m=10 \, M_J$ with au the astronomical unit and $M_J$ the Jupiter mass. Tilted dashed lines show the orbital configurations associated with surface magnetic fields of 5 Gauss and 100 Gauss. Circle (resp. square) point shows the location of Vega with an orbital companion characterised by $m=1.24 \, M_J, D=0.0165 \, \text{au}$ and $P_\text{orb} = 0.53 \, \text{d}$ (resp. $m=0.34 \,  M_J, D=0.017 \, \text{au}$ and $P_\text{orb} = 0.56 \, \text{d}$) as proposed by \citet{boehm2015discovery}.}
	\label{fig:extrapolation_B}
\end{figure*}

\subsubsection*{Comparison with convective dynamo scaling laws}
In planetary or stellar convective dynamos, the viscous dissipation is expected to be negligible compared to the Ohmic one in the limit $Pm \ll 1$, as expected from turbulence studies \citep[e.g.][]{brandenburg2011nonlinear}. In this limit, \citet{davidson2013scaling} argues that dynamo fields should be governed by
\begin{equation}
	B_\text{rms} \sim \sqrt{ \rho_* \, \mu_0}  \, \, (R_* \,  \mathcal{P})^{1/3} ,
	\label{eq:scalingBOD}
\end{equation}
where $\mathcal{P}$ is the power per unit mass injected into the dynamo capable flow (i.e. convection for convective dynamos, and tidal instability here). In this limit of vanishing viscous dissipation, it turns out that equation (\ref{eq:scalingBOD}) is also consistent with the scaling laws obtained from usual convective dynamo simulations using the Boussinesq approximation \citep[e.g.][]{schrinner2012dipole,yadav2013consistent,oruba2014predictive}. The power law given by equation (\ref{eq:scalingBOD}) also holds for anelastic convection \citep{yadav2013scaling}. Thus, we can rely on equation (\ref{eq:scalingBOD}) to compare convective dynamos scaling laws with our empirical scaling law (\ref{eq:scalingB}). 

To estimate $\mathcal{P}$, one can consider the tidal instability in a regime where viscous and ohmic dissipations are of the same order, such that any scaling law obtained for the viscous (or the ohmic) dissipation would also govern $\mathcal{P}$. This regime has been numerically studied by \cite{barker2013nonb} by imposing a weak magnetic field in a periodic box. They obtained that the dissipation rate per unit mass $\mathcal{D}_\nu$ is given by
\begin{equation}
\mathcal{D}_\nu=\chi (2 R_*)^2 |\Omega_s - \Omega_\text{orb}|^3 \epsilon_*^3, \label{eq:Dnu}
\end{equation}
with $\chi \simeq 10^{-2}$. 
Hence, assuming $\mathcal{P} \sim \mathcal{D}_\nu$, equations (\ref{eq:scalingBOD}) and (\ref{eq:Dnu}) give the surface magnetic field $B_0=f_1^{1/2} B_\text{rms}$ as
\begin{equation}
	B_0  \sim \delta \, \epsilon_* \sqrt{ \rho_* \, \mu_0}  \, \, |\Omega_s - \Omega_\text{orb}| R_* ,
	\label{eq:scalingBOD2}
\end{equation}
with $\delta=\sqrt{f_1 f_2}$ and $f_2= (4 \chi)^{2/3} \approx 0.1$. Thus, we recover equation (\ref{eq:scalingB}) exactly. Moreover, \cite{aubert2017spherical} obtained $f_1 \approx 10^{-2}$ for a set of (convective) geodynamo simulations. Therefore, the scaling laws proposed for convective dynamo simulations are fully consistent with our scaling law (\ref{eq:scalingB}), with a similar prefactor $\delta \in [10^{-3},10^{-2}]$.
This gives some confidence in the extrapolations to stars that follows.

We show in figure \ref{fig:extrapolation_B} (a) the growth rate given by formula (\ref{eq:sigma_TDEI2}) and in (b) the surface field strength given by formula (\ref{eq:scalingB}), for various orbital configurations. We have separated physical quantities of the orbital companion, (shown on the vertical axis) from stellar parameters (shown on the horizontal axis).
Assuming that a Hot Jupiter is orbiting around the host star ($m \leq 10 \, M_J$ with $M_J$ the Jupiter mass and $D \geq 0.01$~au), we expect magnetic field strengths ranging from sub-Gauss values to thousands of Gauss.
Thus, tidal dynamos cannot be discarded in tidally deformed radiative stars with moderate stratification ($N_0/\Omega_s \lesssim 2-10$).

\subsubsection*{Tidally driven dynamos in Vega-like stars?}

Vega, with mass $M_* = 2.15 \, M_\odot$, radius $R_* = 2.5 R_\odot$, period $P_s = 0.68$~d \citep{alina2012long,boehm2015discovery} has a surface field strength of order $B_0 = 0.6 \pm 0.3 $~G \citep{lignieres2009first,petit2010rapid}.
The fossil field theory predicts a field strength  $B_0 = 20$ G \citep{braithwaite2017magnetic}, 20 to 30 times too strong.
To circumvent this issue, \citet{braithwaite2012weak} proposed that Vega contains a non-equilibrium fossil field undergoing dynamic evolution.
Here, we provide an alternative scenario based on tidal forcing.
Indeed, the recent discovery of starspots on Vega \citep{boehm2015discovery} seem to support the existence of a close-in orbiting exoplanet.
An exoplanet with a mass $m=1.24 \, M_J$, at a distance $D=0.0165 \, \text{au}$ from the star, and with an orbital period $P_\text{orb} = 0.53$~d or with a mass $m=0.34 \, M_J$, at distance $D=0.017$~au and with orbital period $P_\text{orb} = 0.56$~d would support the astronomical observations \citep{boehm2015discovery}.
With these parameters, the tidal instability would grow in a few thousands years for the two possible orbital configurations and would yield field strengths of $B_0 \simeq 8$~G for the first planetary configuration or $B_0 \simeq 1.5$~G for the second one, even though the system is close to synchronisation.
Although this requires a moderate stratification, in the lower range of estimated values for Vega \citep[$1\leq N_0/\Omega_s\leq25$ according to][]{rieutord2006dynamics},
the tidal dynamo model is consistent with the observed magnetic field of Vega.
Therefore, Vega-like magnetism could well be due to tidally driven dynamos in tidally deformed bodies.
Moreover, \citet{petit2017spot} suggested that the time dependence of spots at the surface of Vega would support zonal flows, as we have observed in our simulations. This might be another hint supporting our tidal mechanism. 

However, the existence of exoplanets around Vega remains controversial. Extended gaps in the debris disks around host stars are often attributed to tidal perturbations by Hot Jupiter planets. But within the current observational limits, no such massive planets have been detected undoubtedly around Vega \citep{su2013asteroid}. Instead, \citet{zheng2017clearing} proposed a 'lone-planet' scenario to account for the observed structure with a single eccentric gas giant, with a mass $m=3 M_J$ and located at the distance $D=75$ au. This hypothetical exoplanet would be too far from Vega to induce strong tidal effects able to sustain a dynamo field.

\subsubsection*{Tidally driven dynamos in Ap/Bp stars?}
Apart from weak Vega-like magnetism, we assess whether our mechanism is relevant to predict the large field strengths of other possibly tidally deformed magnetic stars, in particular Ap/Bp stars.
Herbig Ae/Be stars, which are the precursors of magnetic Ap/Bp stars in the PMS phase, host magnetic fields with similar configurations than their MS counterparts  \citep{alecian2012high,hubrig2014magnetic}. Hence, it is believed that MS fields of Ap/Bp stars are already present at the PMS phase. About 70~\% of the Herbig Ae/Be stars appear in binary/multiple systems \citep{baines2006binarity}, making them \textit{a priori} good candidates for tidal dynamos. For instance HD 200775 is known to be a non-synchronised binary system. The primary has a dipolar field strength of $1000 \pm 150$~G \citep{alecian2008characterization}.
Yet, the tidal mechanism is unlikely to explain the observed magnetic field, because its intensity predicted using the characteristics of the binary system would be too weak from equation (\ref{eq:scalingB2}).
Indeed, the distance $D$ between the star and its companion is too large to induce strong tidal effects (orbital period of the companion is $P_\text{orb} = 1412$~d and $D = 6.7$~au).

\subsubsection*{Tidal mixing}
The relevance of the fossil field model is well established in chemically peculiar A/B stars \citep[e.g.][]{braithwaite2017magnetic}, in which an in situ magnetic generation by tides is not compatible with our findings.
However, it is worth noting that it does not preclude the existence of the tidal instability within these bodies, in which it could play a dynamical role (without dynamo action).
Indeed, \citet{kama2015fingerprints} suggest that giant planets of mass $m \simeq 0.1-10 \, M_J$ are hiding in at least 30~\% of Herbig Ae/Be disks, possibly inducing strong tidal effects once on the MS (at least for the closest and most massive companions).
Be stars are very rapidly rotating main sequence B star, such as HR 7355 \citep{oksala2010discovery,rivinius2012basic} and HR 5907 \citep{grunhut2011hr}.
Most massive stars ($M_* \geq 8 \, M_\odot$) either are binaries (about 75~\%) or were so at some point in their evolution \citep{sana2012binary}.
Binarity is also a common feature in Be stars \citep{rivinius2013classical}.
Coupled with their rapid rotation periods, typically 0.5 d for HR 7355 \citep{oksala2010discovery,rivinius2012basic}, the tidal instability could be significant in these binary systems (if they are not yet synchronised and if their stratification is not too strong).

\section{Conclusion}
\label{sec:ccl}
	\subsection{Summary}
We have numerically investigated the nonlinear outcome of the tidal instability and assessed its dynamo capability in stellar radiative zones. We have adopted a simplified global model of the equilibrium tide in spherical containers. Its simplicity permits high-resolution numerical simulations using an efficient spectral code \citep{schaeffer2013efficient,schaeffer2017geodynamo}.

We confirm that the basic equilibrium tide is prone to the tidal instability as reported by \citet{cebron2010tidal}.
Furthermore, we have shown that this tidal instability is immune to a stable stratification as long as $N_0/\Omega_s \lesssim 1$.
In non-synchronised bodies the instability grows on the typical time scale $\epsilon_*^{-1} /|\Omega_s - \Omega_\text{orb}|$, yielding typically My for a star with a one day spin period.
The tidal instability induces nonlinear motions, whose typical amplitude scales as $\epsilon_* |\Omega_s -\Omega_\text{orb}| R_*$ \citep{barker2013nonb,barker2016nonb}, regardless of the stratification strength.
These motions can induce radial mixing leading to self-consistent dynamos.

Time-averaged magnetic fields in our dynamos are mostly dipolar, an essential feature for their possible observations by astronomers. 
The dipolar field intensity at the surface is a small fraction $\delta$ of the magnetic intensity in the bulk.
With our proof-of-concept simulations we show that a tidal dynamo is a possible alternative mechanism to explain stellar magnetism of hot intermediate-mass and massive stars hosting close-in orbital companions.

Although motion amplitude being almost independent of the stratification, dynamo action was not found when the stratification is too large.
Provided motion amplitude is large enough so that induction overcomes Ohmic dissipation ($Rm \gtrsim 3000$) and assuming the transitions between regimes occur at values of $N_0/\Omega_s$ independent of the diffusivities, tidally driven dynamos are likely when $N_0/\Omega_s \leq 10$.

By extrapolating our results, we predict (i) a field strength up to several Gauss for presumably realistic orbital configurations (depending on the properties of the orbital companion, such as mass, distance to the host star), (ii) essentially all tidally deformed non-synchronised stars should have fields of strength at least comparable to Vega-like fields. Consequently, tidal dynamos in tidally deformed Vega-like stars could explain their magnetism, provided that they host a large and close enough companion and that their stratification is not too strong ($N_0/\Omega_s \lesssim 2-10$ according to our simulations).
Note also that all proposed mechanisms (e.g. failed fossil fields or innermost convective dynamos) are not mutually exclusive and may be combined to explain the observed fields.

	\subsection{Perspectives}
Our proof-of-concept tidally driven dynamos call for many further studies, both to expand the surveyed 
parameter space and to refine the model.
A considerable amount of work remains to be done to improve direct numerical simulations of tidal flows in stellar interiors, but we already hint at possible astrophysical consequences.

\subsubsection*{Parameter space exploration}

We have not strived to adjust the dimensionless parameters to astrophysical realistic ones in the simulations.
They are out of reach with the numerical resources currently available.
The Reynolds number in well-mixed stars is expected to be huge and only the large scale components of the flow can be simulated. Consequently, the relatively high viscosity regime considered in our simulations may have filtered out tidal instabilities of smaller scales than those already obtained.
We however expect that our proof-of-concept simulations capture the dominant global scales of tidally driven nonlinear motions.
We presume them not to be strongly dependent on resolving much smaller scales, but this is difficult to test numerically.
Further simulations in the low diffusive regime, i.e. $Ek \to 0, Pr\ll1$ and $ Pm\ll1$ are of interest, to study the robustness of tidally driven mixing and dynamo action.
In particular, the dynamo capability in the region $1<N_0/\Omega_s<10$ must be studied with lower diffusivities.
Indeed, the higher critical deformation for the onset of instability (see appendix \ref{sec:append_weakening}) is intriguing and prevents our current simulations to reliably assess the dynamo action in this range.
It would be also worth to infer reliable scaling laws as diffusivities are lowered, especially the behaviour of $\delta$ with $\epsilon_*$ and $Ek$.

Stellar interiors have presumably small Prandtl numbers $10^{-8} \leq Pr \ll 1$ \citep[e.g.][]{rieutord2006dynamics}.
However, we have shown that some mixing is driven by the tidal instability.
Mixed envelopes are often modelled by the assumption of equal turbulent diffusivities, yielding $Pr \lesssim 1$ \citep{zahn1992circulation}. 
The sensitivity of the growth rate with $Pr$ is briefly outlined in appendix \ref{sec:append_weakening} at $Ek=10^{-4}$. 
The dependence on $Pr$ should be better assessed in the future.

\subsubsection*{Model refinements}

Anelastic models of stably stratified stars should be considered to better take into account buoyancy effects \citep{zahn2007magnetic,simitev2017baroclinically}. 
Note that the baroclinic instability has been ruled out from our model. Baroclinic instability is believed to occur in stars \citep{spruit1984baroclinic,kitchatinov2013baroclinic,kitchatinov2014baroclinic}. Only our basic state is barotropic, while the motions driven by the tidal instability are baroclinic. A baroclinic basic state is known to enhance the tidal instability in the equatorial plane of the star \citep{kerswell1993elliptical,le2006thermo}. Moreover, baroclinic basic states generate nonlinear motions which are also dynamo capable, as numerically shown by \citet{simitev2017baroclinically}. Consequently, a baroclinic tidal basic state could be even more dynamo capable and deserves further studies.

The influence of a more realistic geometry is also of interest. Indeed, we have assessed the dynamo capability in the simplest possible geometry of a full container. When a solid inner core is present, the tidal instability is also triggered in ellipsoidal shells \citep{cebron2012elliptical}.
It is known that the global pattern of inertial modes is different in shells  \citep{rieutord1997inertial,dintrans1999gravito,rieutord2010viscous,favier2014non}, which may affect the nonlinear outcome of the tidal instability and ultimately its dynamo capability. However, first numerical \citep{cebron2010systematic,cebron2010tidal} and experimental studies \citep{seyed2004elliptical,lemasquerierlibration} in shells seem in agreement with results obtained in full containers. 

\subsubsection*{Possible astrophysical implications}

Statistically, it is believed that many magnetic stars host yet to be observed companions.
If the tidal instability is responsible for stellar magnetic fields, then our mechanism provides constraints on the companion (e.g. mass, distance).
Further astronomical observations should be carried out to clarify this point, by seeking signatures of orbital planetary companions (star-planets interactions) around magnetic stars or magnetic binaries (star-star interactions). Addressing the relevance of star-star interactions for magnetism of hot stars is one of the objectives of the BinaMIcS collaboration \citep{mathis2013roadmap,alecian2014binamics}.

Then, interactions of the tidal instability with imposed fossil fields need also to be addressed. 
Even in the low $Rm$ limit, in which dynamo action does not occur (if $Rm \leq Rm_c$), the tidal instability could develop against the stabilising effect of the magnetic field in some stars and enhance the Ohmic dissipation of the fossil field due to the tidal mixing. Indeed, star-star interactions may explain that the magnetic incidence is much lower in binaries (less than 1.5 \%) than in isolated stars (around 7\%), as for instance studied by the BinaMIcS collaboration \citep{alecian2014binamics,alecian2017fossil}. Additionally, the time variability induced by the tidal instability may provide an alternative explanation for the observed temporal variability of strong fossil fields in Herbig Ae/Be stars, for instance in HD 190073 \citep{alecian2013dramatic}.

Thus, there is an increasing need for stellar evolution models taking into account mixing in stellar radiative zones, which are often assumed to be motionless \citep{kippenhahn1990stellar}. This assumption is not justified because it does not account for various observational data \citep[e.g.][]{pinsonneault1997mixing}.
Mixing has a strong impact on stellar evolution, for instance injecting hydrogen-rich material in the nuclear core or being responsible for the overabundance of some chemical elements at the surface of massive stars \citep[e.g.][]{maeder2000evolution}.
Various mechanisms have been proposed to account for the observed mixing, such as rotational mixing \citep{zahn1992circulation,zahn2008instabilities}.
Inertia-gravity waves could also partially account for the observed mixing \citep{press1981radiative,garcia1991li,rogers2013internal}.
Inertia-gravity waves propagate in magnetic stars \citep[e.g.][]{neiner2012stochastic} and can be excited by tidal forcing through direct resonances \citep[e.g.][]{dintrans1999gravito,mirouh2016gravito} or parametric resonances (as studied here).
Mixing induced by the tidal instability has been so far overlooked in the models.
However, we have shown that the tidal instability could lead to mixing in stably stratified fluids. Future studies should better quantify the tidal dissipation and mixing efficiency in radiative envelopes to improve future models of stellar evolution.  

\section*{Acknowledgements}
The XSHELLS and SHTns codes are freely available at \url{https://bitbucket.org/nschaeff/}.
The authors thank S. Labrosse (ENS Lyon), B. Favier and M. Le Bars (IRPHE, Marseille) for illuminating discussions about the physical interpretation of numerical results.
JV and DC would like to thank A. ud-Doula (Penn State Worhthington Scranton, Dunmore) for fruitful discussions and very helpful comments.
DC and NS also thank F. Ligni\`eres and L. Jouve (IRAP, Toulouse) for fruitful discussions at the meeting of the ANR IMAGINE (\url{http://userpages.irap.omp.eu/~flignieres/index.html}).
This study was motivated by the BinaMIcS collaboration \citep{mathis2013roadmap,alecian2014binamics,alecian2017fossil} and discussions with S. Mathis (CEA, Paris-Saclay), E. Alecian (IPAG, Toulouse).
JV acknowledges the French {\it Minist\`ere de l'Enseignement Sup\'erieur et de la Recherche} for his Ph.D. grant.
The visit of JV at Univ. of Leeds was supported by Labex OSUG@2020 (ANR10 LABX56) and by the doctoral school TUE of Univ. Grenoble Alpes.
We acknowledge GENCI for awarding us access to resource Occigen (CINES) under grant x2016047382 and x2017047382.
Parts of the computations were also performed on the Froggy platform of CIMENT (\texttt{https://ciment.ujf-grenoble.fr}), supported by the Rh\^ one-Alpes region (CPER07\_13 CIRA), OSUG@2020 LabEx (ANR10 LABX56) and Equip@Meso (ANR10 EQPX-29-01).
This work was partially funded by the French {\it Agence Nationale de la Recherche} under grant ANR-14-CE33-0012 (MagLune) and by the 2017 TelluS program from CNRS-INSU (PNP) AO2017-1040353.
ISTerre is part of Labex OSUG@2020 (ANR10 LABX56).
Most figures were produced using matplotlib (\url{http://matplotlib.org/}) or paraview (\url{https://www.paraview.org/}).




\bibliographystyle{mnras}
\bibliography{bib_mnras} 

\begin{thebibliography}{}
\makeatletter
\relax
\def\mn@urlcharsother{\let\do\@makeother \do\$\do\&\do\#\do\^\do\_\do\%\do\~}
\def\mn@doi{\begingroup\mn@urlcharsother \@ifnextchar [ {\mn@doi@}
  {\mn@doi@[]}}
\def\mn@doi@[#1]#2{\def\@tempa{#1}\ifx\@tempa\@empty \href
  {http://dx.doi.org/#2} {doi:#2}\else \href {http://dx.doi.org/#2} {#1}\fi
  \endgroup}
\def\mn@eprint#1#2{\mn@eprint@#1:#2::\@nil}
\def\mn@eprint@arXiv#1{\href {http://arxiv.org/abs/#1} {{\tt arXiv:#1}}}
\def\mn@eprint@dblp#1{\href {http://dblp.uni-trier.de/rec/bibtex/#1.xml}
  {dblp:#1}}
\def\mn@eprint@#1:#2:#3:#4\@nil{\def\@tempa {#1}\def\@tempb {#2}\def\@tempc
  {#3}\ifx \@tempc \@empty \let \@tempc \@tempb \let \@tempb \@tempa \fi \ifx
  \@tempb \@empty \def\@tempb {arXiv}\fi \@ifundefined
  {mn@eprint@\@tempb}{\@tempb:\@tempc}{\expandafter \expandafter \csname
  mn@eprint@\@tempb\endcsname \expandafter{\@tempc}}}

\bibitem[\protect\citeauthoryear{Akg{\"u}n, Reisenegger, Mastrano  \&
  Marchant}{Akg{\"u}n et~al.}{2013}]{akgun2013stability}
Akg{\"u}n T.,  Reisenegger A.,  Mastrano A.,   Marchant P.,  2013, Monthly
  Notices of the Royal Astronomical Society, 433, 2445

\bibitem[\protect\citeauthoryear{Alecian et~al.,}{Alecian
  et~al.}{2008}]{alecian2008characterization}
Alecian E.,  et~al., 2008, Monthly Notices of the Royal Astronomical Society,
  385, 391

\bibitem[\protect\citeauthoryear{Alecian et~al.,}{Alecian
  et~al.}{2012}]{alecian2012high}
Alecian E.,  et~al., 2012, Monthly Notices of the Royal Astronomical Society,
  429, 1001

\bibitem[\protect\citeauthoryear{Alecian, Neiner, Mathis, Catala, Kochukhov  \&
  Landstreet}{Alecian et~al.}{2013}]{alecian2013dramatic}
Alecian E.,  Neiner C.,  Mathis S.,  Catala C.,  Kochukhov O.,   Landstreet J.,
   2013, Astronomy \& Astrophysics, 549, L8

\bibitem[\protect\citeauthoryear{Alecian et~al.,}{Alecian
  et~al.}{2014}]{alecian2014binamics}
Alecian E.,  et~al., 2014, Proceedings of the International Astronomical Union,
  9, 330

\bibitem[\protect\citeauthoryear{Alecian, Villebrun, Grunhut, Hussain, Neiner
  \& Wade}{Alecian et~al.}{2017}]{alecian2017fossil}
Alecian E.,  Villebrun F.,  Grunhut J.,  Hussain G.,  Neiner C.,   Wade G.~A.,
  2017, arXiv preprint arXiv:1705.10650

\bibitem[\protect\citeauthoryear{Alina, Petit, Ligni{\`e}res, Wade, Fares,
  Auri{\`e}re, B{\"o}hm  \& Carfantan}{Alina et~al.}{2012}]{alina2012long}
Alina D.,  Petit P.,  Ligni{\`e}res F.,  Wade G.,  Fares R.,  Auri{\`e}re M.,
  B{\"o}hm T.,   Carfantan H.,  2012, in AIP Conference Proceedings. pp 82--85

\bibitem[\protect\citeauthoryear{Arlt \& R{\"u}diger}{Arlt \&
  R{\"u}diger}{2011a}]{arlt2011magnetic}
Arlt R.,  R{\"u}diger G.,  2011a, Astronomische Nachrichten, 332, 70

\bibitem[\protect\citeauthoryear{Arlt \& R{\"u}diger}{Arlt \&
  R{\"u}diger}{2011b}]{arlt2011amplification}
Arlt R.,  R{\"u}diger G.,  2011b, Monthly Notices of the Royal Astronomical
  Society, 412, 107

\bibitem[\protect\citeauthoryear{Arlt, Hollerbach  \& R{\"u}diger}{Arlt
  et~al.}{2003}]{arlt2003differential}
Arlt R.,  Hollerbach R.,   R{\"u}diger G.,  2003, Astronomy \& Astrophysics,
  401, 1087

\bibitem[\protect\citeauthoryear{Aubert, Gastine  \& Fournier}{Aubert
  et~al.}{2017}]{aubert2017spherical}
Aubert J.,  Gastine T.,   Fournier A.,  2017, Journal of Fluid Mechanics, 813,
  558

\bibitem[\protect\citeauthoryear{Augustson, Mathis  \& Brun}{Augustson
  et~al.}{2017}]{augustson2016dynamo}
Augustson K.,  Mathis S.,   Brun A.~S.,  2017, arXiv preprint arXiv:1701.02582

\bibitem[\protect\citeauthoryear{Auri{\`e}re et~al.,}{Auri{\`e}re
  et~al.}{2007}]{auriere2007weak}
Auri{\`e}re M.,  et~al., 2007, Astronomy \& Astrophysics, 475, 1053

\bibitem[\protect\citeauthoryear{Babcock}{Babcock}{1947}]{babcock1947zeeman}
Babcock H.~W.,  1947, The Astrophysical Journal, 105, 105

\bibitem[\protect\citeauthoryear{Baines, Oudmaijer, Porter  \& Pozzo}{Baines
  et~al.}{2006}]{baines2006binarity}
Baines D.,  Oudmaijer R.~D.,  Porter J.~M.,   Pozzo M.,  2006, Monthly Notices
  of the Royal Astronomical Society, 367, 737

\bibitem[\protect\citeauthoryear{Balbus \& Hawley}{Balbus \&
  Hawley}{1991}]{balbus1991powerful}
Balbus S.~A.,  Hawley J.~F.,  1991, The Astrophysical Journal, 376, 214

\bibitem[\protect\citeauthoryear{Barker}{Barker}{2016}]{barker2016nonb}
Barker A.~J.,  2016, Monthly Notices of the Royal Astronomical Society, 459,
  939

\bibitem[\protect\citeauthoryear{Barker \& Lithwick}{Barker \&
  Lithwick}{2013a}]{barker2013nona}
Barker A.~J.,  Lithwick Y.,  2013a, Monthly Notices of the Royal Astronomical
  Society, 435, 3614

\bibitem[\protect\citeauthoryear{Barker \& Lithwick}{Barker \&
  Lithwick}{2013b}]{barker2013nonb}
Barker A.~J.,  Lithwick Y.,  2013b, Monthly Notices of the Royal Astronomical
  Society, 437, 305

\bibitem[\protect\citeauthoryear{Barker, Braviner  \& Ogilvie}{Barker
  et~al.}{2016}]{barker2016non}
Barker A.~J.,  Braviner H.~J.,   Ogilvie G.~I.,  2016, Monthly Notices of the
  Royal Astronomical Society, 459, 924

\bibitem[\protect\citeauthoryear{Behrend \& Maeder}{Behrend \&
  Maeder}{2001}]{behrend2001formation}
Behrend R.,  Maeder A.,  2001, Astronomy \& Astrophysics, 373, 190

\bibitem[\protect\citeauthoryear{Blaz{\`e}re, Neiner  \& Petit}{Blaz{\`e}re
  et~al.}{2016a}]{blazere2016discovery}
Blaz{\`e}re A.,  Neiner C.,   Petit P.,  2016a, Monthly Notices of the Royal
  Astronomical Society: Letters, 459, L81

\bibitem[\protect\citeauthoryear{Blaz{\`e}re et~al.,}{Blaz{\`e}re
  et~al.}{2016b}]{blazere2016detection}
Blaz{\`e}re A.,  et~al., 2016b, Astronomy \& Astrophysics, 586, A97

\bibitem[\protect\citeauthoryear{Boehm et~al.,}{Boehm
  et~al.}{2015}]{boehm2015discovery}
Boehm T.,  et~al., 2015, Astronomy \& Astrophysics, 577, A64

\bibitem[\protect\citeauthoryear{Borra, Landstreet  \& Mestel}{Borra
  et~al.}{1982}]{borra1982magnetic}
Borra E.~F.,  Landstreet J.~D.,   Mestel L.,  1982, Annual Review of Astronomy
  and Astrophysics, 20, 191

\bibitem[\protect\citeauthoryear{Braithwaite}{Braithwaite}{2006}]{braithwaite2006differential}
Braithwaite J.,  2006, Astronomy \& Astrophysics, 449, 451

\bibitem[\protect\citeauthoryear{Braithwaite \& Cantiello}{Braithwaite \&
  Cantiello}{2012}]{braithwaite2012weak}
Braithwaite J.,  Cantiello M.,  2012, Monthly Notices of the Royal Astronomical
  Society, 428, 2789

\bibitem[\protect\citeauthoryear{Braithwaite \& Spruit}{Braithwaite \&
  Spruit}{2004}]{braithwaite2004fossil}
Braithwaite J.,  Spruit H.~C.,  2004, Nature, 431, 819

\bibitem[\protect\citeauthoryear{Braithwaite \& Spruit}{Braithwaite \&
  Spruit}{2017}]{braithwaite2017magnetic}
Braithwaite J.,  Spruit H.~C.,  2017, Royal Society Open Science, 4, 160271

\bibitem[\protect\citeauthoryear{Brandenburg}{Brandenburg}{2001}]{brandenburg2001inverse}
Brandenburg A.,  2001, The Astrophysical Journal, 550, 824

\bibitem[\protect\citeauthoryear{Brandenburg}{Brandenburg}{2009}]{brandenburg2009large}
Brandenburg A.,  2009, The Astrophysical Journal, 697, 1206

\bibitem[\protect\citeauthoryear{Brandenburg}{Brandenburg}{2011}]{brandenburg2011nonlinear}
Brandenburg A.,  2011, The Astrophysical Journal, 741, 92

\bibitem[\protect\citeauthoryear{Brun, Miesch  \& Toomre}{Brun
  et~al.}{2004}]{brun2004global}
Brun A.~S.,  Miesch M.~S.,   Toomre J.,  2004, The Astrophysical Journal, 614,
  1073

\bibitem[\protect\citeauthoryear{Bullard \& Gellman}{Bullard \&
  Gellman}{1954}]{bullard1954homogeneous}
Bullard E.,  Gellman H.,  1954, Philosophical Transactions of the Royal Society
  of London. Series A, Mathematical and Physical Sciences, pp 213--278

\bibitem[\protect\citeauthoryear{Busse}{Busse}{1970}]{busse1970thermal}
Busse F.~H.,  1970, Journal of Fluid Mechanics, 44, 441

\bibitem[\protect\citeauthoryear{Buysschaert, Neiner, Briquet  \&
  Aerts}{Buysschaert et~al.}{2017}]{buysschaert2017magnetic}
Buysschaert B.,  Neiner C.,  Briquet M.,   Aerts C.,  2017, Astronomy \&
  Astrophysics, 605, A104

\bibitem[\protect\citeauthoryear{Cantiello \& Braithwaite}{Cantiello \&
  Braithwaite}{2011}]{cantiello2011magnetic}
Cantiello M.,  Braithwaite J.,  2011, Astronomy \& Astrophysics, 534, A140

\bibitem[\protect\citeauthoryear{C{\'e}bron \& Hollerbach}{C{\'e}bron \&
  Hollerbach}{2014}]{cebron2014tidally}
C{\'e}bron D.,  Hollerbach R.,  2014, The Astrophysical Journal Letters, 789,
  L25

\bibitem[\protect\citeauthoryear{C{\'e}bron, Le~Bars, Leontini, Maubert  \&
  Le~Gal}{C{\'e}bron et~al.}{2010a}]{cebron2010systematic}
C{\'e}bron D.,  Le~Bars M.,  Leontini J.,  Maubert P.,   Le~Gal P.,  2010a,
  Physics of the Earth and Planetary Interiors, 182, 119

\bibitem[\protect\citeauthoryear{C{\'e}bron, Maubert  \& Le~Bars}{C{\'e}bron
  et~al.}{2010b}]{cebron2010tidal}
C{\'e}bron D.,  Maubert P.,   Le~Bars M.,  2010b, Geophysical Journal
  International, 182, 1311

\bibitem[\protect\citeauthoryear{C{\'e}bron, Le~Bars, Maubert  \&
  Le~Gal}{C{\'e}bron et~al.}{2012a}]{cebron2012magnetohydrodynamic}
C{\'e}bron D.,  Le~Bars M.,  Maubert P.,   Le~Gal P.,  2012a, Geophysical \&
  Astrophysical Fluid Dynamics, 106, 524

\bibitem[\protect\citeauthoryear{C{\'e}bron, Le~Bars, Moutou  \&
  Le~Gal}{C{\'e}bron et~al.}{2012b}]{cebron2012elliptical}
C{\'e}bron D.,  Le~Bars M.,  Moutou C.,   Le~Gal P.,  2012b, Astronomy \&
  Astrophysics, 539, A78

\bibitem[\protect\citeauthoryear{C{\'e}bron, Le~Bars, Le~Gal, Moutou, Leconte
  \& Sauret}{C{\'e}bron et~al.}{2013}]{cebron2013elliptical}
C{\'e}bron D.,  Le~Bars M.,  Le~Gal P.,  Moutou C.,  Leconte J.,   Sauret A.,
  2013, Icarus, 226, 1642

\bibitem[\protect\citeauthoryear{Charbonneau}{Charbonneau}{2014}]{charbonneau2014solar}
Charbonneau P.,  2014, Annual Review of Astronomy and Astrophysics, 52, 251

\bibitem[\protect\citeauthoryear{Charbonneau \& MacGregor}{Charbonneau \&
  MacGregor}{2001}]{charbonneau2001magnetic}
Charbonneau P.,  MacGregor K.~B.,  2001, The Astrophysical Journal, 559, 1094

\bibitem[\protect\citeauthoryear{Christensen, Holzwarth  \&
  Reiners}{Christensen et~al.}{2009}]{christensen2009energy}
Christensen U.~R.,  Holzwarth V.,   Reiners A.,  2009, Nature, 457, 167

\bibitem[\protect\citeauthoryear{Davidson}{Davidson}{2013}]{davidson2013scaling}
Davidson P.,  2013, Geophysical Journal International, 195, 67

\bibitem[\protect\citeauthoryear{Dintrans, Rieutord  \& Valdettaro}{Dintrans
  et~al.}{1999}]{dintrans1999gravito}
Dintrans B.,  Rieutord M.,   Valdettaro L.,  1999, Journal of Fluid Mechanics,
  398, 271

\bibitem[\protect\citeauthoryear{Donati \& Landstreet}{Donati \&
  Landstreet}{2009}]{donati2009magnetic}
Donati J.-F.,  Landstreet J.~D.,  2009, Annual Review of Astronomy and
  Astrophysics, 47, 333

\bibitem[\protect\citeauthoryear{Fares et~al.,}{Fares
  et~al.}{2009}]{fares2009magnetic}
Fares R.,  et~al., 2009, Monthly Notices of the Royal Astronomical Society,
  398, 1383

\bibitem[\protect\citeauthoryear{Fares et~al.,}{Fares
  et~al.}{2012}]{fares2012magnetic}
Fares R.,  et~al., 2012, Monthly Notices of the Royal Astronomical Society,
  423, 1006

\bibitem[\protect\citeauthoryear{Fares et~al.,}{Fares
  et~al.}{2017}]{fares2017moves}
Fares R.,  et~al., 2017, Monthly Notices of the Royal Astronomical Society,
  471, 1246

\bibitem[\protect\citeauthoryear{Favier, Barker, Baruteau  \& Ogilvie}{Favier
  et~al.}{2014}]{favier2014non}
Favier B.,  Barker A.~J.,  Baruteau C.,   Ogilvie G.~I.,  2014, Monthly Notices
  of the Royal Astronomical Society, 439, 845

\bibitem[\protect\citeauthoryear{Favier, Grannan, Le~Bars  \& Aurnou}{Favier
  et~al.}{2015}]{favier2015generation}
Favier B.,  Grannan A.~M.,  Le~Bars M.,   Aurnou J.~M.,  2015, Physics of
  Fluids, 27, 066601

\bibitem[\protect\citeauthoryear{Featherstone, Browning, Brun  \&
  Toomre}{Featherstone et~al.}{2009}]{featherstone2009effects}
Featherstone N.~A.,  Browning M.~K.,  Brun A.~S.,   Toomre J.,  2009, The
  Astrophysical Journal, 705, 1000

\bibitem[\protect\citeauthoryear{Ferriz-Mas, Schmitt  \&
  Sch{\"u}ssler}{Ferriz-Mas et~al.}{1994}]{ferriz1994dynamo}
Ferriz-Mas A.,  Schmitt D.,   Sch{\"u}ssler M.,  1994, Astronomy and
  Astrophysics, 289, 949

\bibitem[\protect\citeauthoryear{Friedlander \& Siegmann}{Friedlander \&
  Siegmann}{1982}]{friedlander1982internala}
Friedlander S.,  Siegmann W.~L.,  1982, Geophysical \& Astrophysical Fluid
  Dynamics, 19, 267

\bibitem[\protect\citeauthoryear{Garcia~Lopez \& Spruit}{Garcia~Lopez \&
  Spruit}{1991}]{garcia1991li}
Garcia~Lopez R.~J.,  Spruit H.~C.,  1991, The Astrophysical Journal, 377, 268

\bibitem[\protect\citeauthoryear{Gellert, R{\"u}diger  \& Hollerbach}{Gellert
  et~al.}{2011}]{gellert2011helicity}
Gellert M.,  R{\"u}diger G.,   Hollerbach R.,  2011, Monthly Notices of the
  Royal Astronomical Society, 414, 2696

\bibitem[\protect\citeauthoryear{Glatzmaier \& Roberts}{Glatzmaier \&
  Roberts}{1995}]{glatzmaiers1995three}
Glatzmaier G.~A.,  Roberts P.~H.,  1995, Nature, 377, 203

\bibitem[\protect\citeauthoryear{Goepfert \& Tilgner}{Goepfert \&
  Tilgner}{2016}]{goepfert2016dynamos}
Goepfert O.,  Tilgner A.,  2016, New Journal of Physics, 18, 103019

\bibitem[\protect\citeauthoryear{Goodman}{Goodman}{1993}]{goodman1993local}
Goodman J.,  1993, The Astrophysical Journal, 406, 596

\bibitem[\protect\citeauthoryear{Grannan, Favier, Le~Bars  \& Aurnou}{Grannan
  et~al.}{2016}]{grannan2016tidally}
Grannan A.~M.,  Favier B.,  Le~Bars M.,   Aurnou J.~M.,  2016, Geophysical
  Journal International, 208, 1690

\bibitem[\protect\citeauthoryear{Grunhut et~al.,}{Grunhut
  et~al.}{2011}]{grunhut2011hr}
Grunhut J.~H.,  et~al., 2011, Monthly Notices of the Royal Astronomical
  Society, 419, 1610

\bibitem[\protect\citeauthoryear{Guervilly \& Cardin}{Guervilly \&
  Cardin}{2010}]{guervilly2010numerical}
Guervilly C.,  Cardin P.,  2010, Geophysical \& Astrophysical Fluid Dynamics,
  104, 221

\bibitem[\protect\citeauthoryear{Guervilly, Hughes  \& Jones}{Guervilly
  et~al.}{2014}]{guervilly2014large}
Guervilly C.,  Hughes D.~W.,   Jones C.~A.,  2014, Journal of Fluid Mechanics,
  758, 407

\bibitem[\protect\citeauthoryear{Guervilly, Hughes  \& Jones}{Guervilly
  et~al.}{2015}]{guervilly2015generation}
Guervilly C.,  Hughes D.~W.,   Jones C.~A.,  2015, Physical Review E, 91,
  041001

\bibitem[\protect\citeauthoryear{Hale}{Hale}{1908}]{hale1908probable}
Hale G.~E.,  1908, The Astrophysical Journal, 28, 315

\bibitem[\protect\citeauthoryear{Herreman, C{\'e}bron, Le~Diz{\`e}s  \&
  Le~Gal}{Herreman et~al.}{2010}]{herreman2010elliptical}
Herreman W.,  C{\'e}bron D.,  Le~Diz{\`e}s S.,   Le~Gal P.,  2010, Journal of
  Fluid Mechanics, 661, 130

\bibitem[\protect\citeauthoryear{Hubrig, Briquet, Sch{\"o}ller, De~Cat, Mathys
  \& Aerts}{Hubrig et~al.}{2006}]{hubrig2006discovery}
Hubrig S.,  Briquet M.,  Sch{\"o}ller M.,  De~Cat P.,  Mathys G.,   Aerts C.,
  2006, Monthly Notices of the Royal Astronomical Society: Letters, 369, L61

\bibitem[\protect\citeauthoryear{Hubrig, Ilyin, Sch{\"o}ller, Cowley, Castelli,
  Stelzer, Gonzalez  \& Wolff}{Hubrig et~al.}{2014}]{hubrig2014magnetic}
Hubrig S.,  Ilyin I.,  Sch{\"o}ller M.,  Cowley C.~R.,  Castelli F.,  Stelzer
  B.,  Gonzalez J.-F.,   Wolff B.,  2014, in EPJ Web of Conferences. p. 08006

\bibitem[\protect\citeauthoryear{Jouve \& Brun}{Jouve \&
  Brun}{2007}]{jouve2007role}
Jouve L.,  Brun A.~S.,  2007, Astronomy \& Astrophysics, 474, 239

\bibitem[\protect\citeauthoryear{Jouve, Brown  \& Brun}{Jouve
  et~al.}{2010}]{jouve2010exploring}
Jouve L.,  Brown B.~P.,   Brun A.~S.,  2010, Astronomy \& Astrophysics, 509,
  A32

\bibitem[\protect\citeauthoryear{Jouve, Gastine  \& Ligni{\`e}res}{Jouve
  et~al.}{2015}]{jouve2015three}
Jouve L.,  Gastine T.,   Ligni{\`e}res F.,  2015, Astronomy \& Astrophysics,
  575, A106

\bibitem[\protect\citeauthoryear{Kaiser \& Busse}{Kaiser \&
  Busse}{2017}]{kaiser2017robustness}
Kaiser R.,  Busse F.,  2017, Geophysical \& Astrophysical Fluid Dynamics, pp
  1--14

\bibitem[\protect\citeauthoryear{Kama, Folsom  \& Pinilla}{Kama
  et~al.}{2015}]{kama2015fingerprints}
Kama M.,  Folsom C.~P.,   Pinilla P.,  2015, Astronomy \& Astrophysics, 582,
  L10

\bibitem[\protect\citeauthoryear{Kerswell}{Kerswell}{1993}]{kerswell1993elliptical}
Kerswell R.~R.,  1993, Geophysical \& Astrophysical Fluid Dynamics, 71, 105

\bibitem[\protect\citeauthoryear{Kerswell}{Kerswell}{2002}]{kerswell2002elliptical}
Kerswell R.~R.,  2002, Annual Review of Fluid Mechanics, 34, 83

\bibitem[\protect\citeauthoryear{Kippenhahn, Weigert  \& Weiss}{Kippenhahn
  et~al.}{1990}]{kippenhahn1990stellar}
Kippenhahn R.,  Weigert A.,   Weiss A.,  1990, Stellar structure and evolution.
Springer

\bibitem[\protect\citeauthoryear{Kitchatinov}{Kitchatinov}{2013}]{kitchatinov2013baroclinic}
Kitchatinov L.~L.,  2013, Astronomy Letters, 39, 561

\bibitem[\protect\citeauthoryear{Kitchatinov}{Kitchatinov}{2014}]{kitchatinov2014baroclinic}
Kitchatinov L.~L.,  2014, The Astrophysical Journal, 784, 81

\bibitem[\protect\citeauthoryear{Kochukhov \& Sudnik}{Kochukhov \&
  Sudnik}{2013}]{kochukhov2013detectability}
Kochukhov O.,  Sudnik N.,  2013, Astronomy \& Astrophysics, 554, A93

\bibitem[\protect\citeauthoryear{Lacaze, Herreman, Le~Bars, Le~Diz{\`e}s  \&
  Le~Gal}{Lacaze et~al.}{2006}]{lacaze2006magnetic}
Lacaze L.,  Herreman W.,  Le~Bars M.,  Le~Diz{\`e}s S.,   Le~Gal P.,  2006,
  Geophysical and Astrophysical Fluid Dynamics, 100, 299

\bibitem[\protect\citeauthoryear{Larmor}{Larmor}{1919}]{larmor1919could}
Larmor J.,  1919, Rep. Brit. Assoc. Adv. Sci, 159, 412

\bibitem[\protect\citeauthoryear{Le~Bars \& Le~Diz{\`e}s}{Le~Bars \&
  Le~Diz{\`e}s}{2006}]{le2006thermo}
Le~Bars M.,  Le~Diz{\`e}s S.,  2006, Journal of Fluid Mechanics, 563, 189

\bibitem[\protect\citeauthoryear{Le~Bars, Lacaze, Le~Diz{\`e}s, Le~Gal  \&
  Rieutord}{Le~Bars et~al.}{2010}]{le2010tidal}
Le~Bars M.,  Lacaze L.,  Le~Diz{\`e}s S.,  Le~Gal P.,   Rieutord M.,  2010,
  Physics of the Earth and Planetary Interiors, 178, 48

\bibitem[\protect\citeauthoryear{Le~Reun, Favier, Barker  \& Le~Bars}{Le~Reun
  et~al.}{2017}]{le2017inertial}
Le~Reun T.,  Favier B.,  Barker A.~J.,   Le~Bars M.,  2017, Physical Review
  Letters, 119, 034502

\bibitem[\protect\citeauthoryear{Lemasquerier, Grannan, Vidal, C{\'e}bron,
  Favier, Le~Bars  \& Aurnou}{Lemasquerier
  et~al.}{2017}]{lemasquerierlibration}
Lemasquerier D.,  Grannan A.~M.,  Vidal J.,  C{\'e}bron D.,  Favier B.,
  Le~Bars M.,   Aurnou J.~M.,  2017, \mn@doi [Journal of Geophysical Research:
  Planets] {10.1002/2017JE005340}

\bibitem[\protect\citeauthoryear{Lewis \& Bellan}{Lewis \&
  Bellan}{1990}]{lewis1990physical}
Lewis H.~R.,  Bellan P.~M.,  1990, Journal of Mathematical Physics, 31, 2592

\bibitem[\protect\citeauthoryear{Ligni{\`e}res, Petit, B{\"o}hm  \&
  Auriere}{Ligni{\`e}res et~al.}{2009}]{lignieres2009first}
Ligni{\`e}res F.,  Petit P.,  B{\"o}hm T.,   Auriere M.,  2009, Astronomy \&
  Astrophysics, 500, L41

\bibitem[\protect\citeauthoryear{Ligni{\`e}res, Petit, Auri{\`e}re, Wade  \&
  B{\"o}hm}{Ligni{\`e}res et~al.}{2013}]{lignieres2013dichotomy}
Ligni{\`e}res F.,  Petit P.,  Auri{\`e}re M.,  Wade G.~A.,   B{\"o}hm T.,
  2013, Proceedings of the International Astronomical Union, 9, 338

\bibitem[\protect\citeauthoryear{Lin \& Ogilvie}{Lin \&
  Ogilvie}{2017}]{lin2017tidal}
Lin Y.,  Ogilvie G.~I.,  2017, Monthly Notices of the Royal Astronomical
  Society, 468, 1387

\bibitem[\protect\citeauthoryear{Livermore, Bailey  \& Hollerbach}{Livermore
  et~al.}{2016}]{livermore2016comparison}
Livermore P.~W.,  Bailey L.~M.,   Hollerbach R.,  2016, Scientific Reports, 6,
  22812

\bibitem[\protect\citeauthoryear{MacDonald \& Mullan}{MacDonald \&
  Mullan}{2004}]{macdonald2004magnetic}
MacDonald J.,  Mullan D.~J.,  2004, Monthly Notices of the Royal Astronomical
  Society, 348, 702

\bibitem[\protect\citeauthoryear{MacGregor \& Cassinelli}{MacGregor \&
  Cassinelli}{2003}]{macgregor2003magnetic}
MacGregor K.~B.,  Cassinelli J.~P.,  2003, The Astrophysical Journal, 586, 480

\bibitem[\protect\citeauthoryear{Maeder \& Meynet}{Maeder \&
  Meynet}{2000}]{maeder2000evolution}
Maeder A.,  Meynet G.,  2000, Annual Review of Astronomy and Astrophysics, 38,
  143

\bibitem[\protect\citeauthoryear{Marcotte \& Gissinger}{Marcotte \&
  Gissinger}{2016}]{marcotte2016dynamo}
Marcotte F.,  Gissinger C.,  2016, Physical Review Fluids, 1, 063602

\bibitem[\protect\citeauthoryear{Markey \& Tayler}{Markey \&
  Tayler}{1973}]{markey1973adiabatic}
Markey P.,  Tayler R.~J.,  1973, Monthly Notices of the Royal Astronomical
  Society, 163, 77

\bibitem[\protect\citeauthoryear{Marti et~al.,}{Marti et~al.}{2014}]{marti2014}
Marti P.,  et~al., 2014, \mn@doi [Geophysical Journal International]
  {10.1093/gji/ggt518}, 197, 119

\bibitem[\protect\citeauthoryear{Mathis, Neiner, Alecian  \& Wade}{Mathis
  et~al.}{2013}]{mathis2013roadmap}
Mathis S.,  Neiner C.,  Alecian E.,   Wade G.,  2013, Proceedings of the
  International Astronomical Union, 9, 311

\bibitem[\protect\citeauthoryear{Mathys}{Mathys}{2017}]{mathys2017ap}
Mathys G.,  2017, Astronomy \& Astrophysics, 601, A14

\bibitem[\protect\citeauthoryear{Matsui et~al.,}{Matsui
  et~al.}{2016}]{matsui2016performance}
Matsui H.,  et~al., 2016, Geochemistry, Geophysics, Geosystems, 17, 1586

\bibitem[\protect\citeauthoryear{Mininni}{Mininni}{2007}]{mininni2007inverse}
Mininni P.~D.,  2007, Physical Review E, 76, 026316

\bibitem[\protect\citeauthoryear{Mininni, Ponty, Montgomery, Pinton, Politano
  \& Pouquet}{Mininni et~al.}{2005}]{mininni2005dynamo}
Mininni P.~D.,  Ponty Y.,  Montgomery D.~C.,  Pinton J.-F.,  Politano H.,
  Pouquet A.,  2005, The Astrophysical Journal, 626, 853

\bibitem[\protect\citeauthoryear{Mirouh, Baruteau, Rieutord  \& Ballot}{Mirouh
  et~al.}{2016}]{mirouh2016gravito}
Mirouh G.~M.,  Baruteau C.,  Rieutord M.,   Ballot J.,  2016, Journal of Fluid
  Mechanics, 800, 213

\bibitem[\protect\citeauthoryear{Miyazaki}{Miyazaki}{1993}]{miyazaki1993elliptical}
Miyazaki T.,  1993, Physics of Fluids A: Fluid Dynamics, 5, 2702

\bibitem[\protect\citeauthoryear{Mizerski \& Lyra}{Mizerski \&
  Lyra}{2012}]{mizerski2012connection}
Mizerski K.~A.,  Lyra W.,  2012, Journal of Fluid Mechanics, 698, 358

\bibitem[\protect\citeauthoryear{Mizerski, Bajer  \& Moffatt}{Mizerski
  et~al.}{2012}]{mizerski2012mean}
Mizerski K.~A.,  Bajer K.,   Moffatt H.~K.,  2012, Journal of Fluid Mechanics,
  707, 111

\bibitem[\protect\citeauthoryear{Moss}{Moss}{1989}]{moss1989origin}
Moss D.,  1989, Monthly Notices of the Royal Astronomical Society, 236, 629

\bibitem[\protect\citeauthoryear{Neiner et~al.,}{Neiner
  et~al.}{2012}]{neiner2012stochastic}
Neiner C.,  et~al., 2012, Astronomy \& Astrophysics, 546, A47

\bibitem[\protect\citeauthoryear{Neiner, Mathis, Alecian, Emeriau  \&
  Grunhut}{Neiner et~al.}{2014}]{neiner2014origin}
Neiner C.,  Mathis S.,  Alecian E.,  Emeriau C.,   Grunhut J.,  2014,
  Proceedings of the International Astronomical Union, 10, 61

\bibitem[\protect\citeauthoryear{Ogilvie}{Ogilvie}{2005}]{ogilvie2005wave}
Ogilvie G.~I.,  2005, Journal of Fluid Mechanics, 543, 19

\bibitem[\protect\citeauthoryear{Ogilvie}{Ogilvie}{2013}]{ogilvie2013tides}
Ogilvie G.~I.,  2013, Monthly Notices of the Royal Astronomical Society, 429,
  613

\bibitem[\protect\citeauthoryear{Oksala, Wade, Marcolino, Grunhut, Bohlender,
  Manset  \& Townsend}{Oksala et~al.}{2010}]{oksala2010discovery}
Oksala M.~E.,  Wade G.~A.,  Marcolino W. L.~F.,  Grunhut J.,  Bohlender D.,
  Manset N.,   Townsend R. H.~D.,  2010, Monthly Notices of the Royal
  Astronomical Society: Letters, 405, L51

\bibitem[\protect\citeauthoryear{Oruba \& Dormy}{Oruba \&
  Dormy}{2014}]{oruba2014predictive}
Oruba L.,  Dormy E.,  2014, Geophysical Journal International, 198, 828

\bibitem[\protect\citeauthoryear{Palla \& Stahler}{Palla \&
  Stahler}{1992}]{palla1992evolution}
Palla F.,  Stahler S.~W.,  1992, The Astrophysical Journal, 392, 667

\bibitem[\protect\citeauthoryear{Parker}{Parker}{1955}]{parker1955hydromagnetic}
Parker E.~N.,  1955, The Astrophysical Journal, 122, 293

\bibitem[\protect\citeauthoryear{Parker}{Parker}{1975}]{parker1975generation}
Parker E.~N.,  1975, The Astrophysical Journal, 198, 205

\bibitem[\protect\citeauthoryear{Parker}{Parker}{1979}]{parker1979cosmical}
Parker E.~N.,  1979, Cosmical magnetic fields: Their origin and their activity.
Clarendon Press; Oxford University Press

\bibitem[\protect\citeauthoryear{Petit et~al.,}{Petit
  et~al.}{2010}]{petit2010rapid}
Petit P.,  et~al., 2010, Astronomy \& Astrophysics, 523, A41

\bibitem[\protect\citeauthoryear{Petit et~al.,}{Petit
  et~al.}{2011}]{petit2011detection}
Petit P.,  et~al., 2011, Astronomy \& Astrophysics, 532, L13

\bibitem[\protect\citeauthoryear{Petit, H{\'e}brard, B{\"o}hm, Folsom  \&
  Ligni{\`e}res}{Petit et~al.}{2017}]{petit2017spot}
Petit P.,  H{\'e}brard E.~M.,  B{\"o}hm T.,  Folsom C.~P.,   Ligni{\`e}res F.,
  2017, Monthly Notices of the Royal Astronomical Society: Letters, 472, L30

\bibitem[\protect\citeauthoryear{Pinsonneault}{Pinsonneault}{1997}]{pinsonneault1997mixing}
Pinsonneault M.,  1997, Annual Review of Astronomy and Astrophysics, 35, 557

\bibitem[\protect\citeauthoryear{Pitts \& Tayler}{Pitts \&
  Tayler}{1985}]{pitts1985adiabatic}
Pitts E.,  Tayler R.~J.,  1985, Monthly Notices of the Royal Astronomical
  Society, 216, 139

\bibitem[\protect\citeauthoryear{Ponty, Politano  \& Pinton}{Ponty
  et~al.}{2004}]{ponty2004simulation}
Ponty Y.,  Politano H.,   Pinton J.-F.,  2004, Physical Review Letters, 92,
  144503

\bibitem[\protect\citeauthoryear{Ponty, Mininni, Montgomery, Pinton, Politano
  \& Pouquet}{Ponty et~al.}{2005}]{ponty2005numerical}
Ponty Y.,  Mininni P.~D.,  Montgomery D.~C.,  Pinton J.-F.,  Politano H.,
  Pouquet A.,  2005, Physical Review Letters, 94, 164502

\bibitem[\protect\citeauthoryear{Ponty, Mininni, Pinton, Politano  \&
  Pouquet}{Ponty et~al.}{2007}]{ponty2007dynamo}
Ponty Y.,  Mininni P.~D.,  Pinton J.-F.,  Politano H.,   Pouquet A.,  2007, New
  Journal of Physics, 9, 296

\bibitem[\protect\citeauthoryear{Potter, Chitre  \& Tout}{Potter
  et~al.}{2012}]{potter2012stellar}
Potter A.~T.,  Chitre S.~M.,   Tout C.~A.,  2012, Monthly Notices of the Royal
  Astronomical Society, 424, 2358

\bibitem[\protect\citeauthoryear{Power, Wade, Auri{\`e}re, Silvester  \&
  Hanes}{Power et~al.}{2008}]{power2008properties}
Power J.,  Wade G.~A.,  Auri{\`e}re M.,  Silvester J.,   Hanes D.,  2008,
  Contrib. Astron. Obs. Skalnat{\'e} Pleso, 38, 443

\bibitem[\protect\citeauthoryear{Press}{Press}{1981}]{press1981radiative}
Press W.~H.,  1981, The Astrophysical Journal, 245, 286

\bibitem[\protect\citeauthoryear{Remus, Mathis  \& Zahn}{Remus
  et~al.}{2012}]{remus2012equilibrium}
Remus F.,  Mathis S.,   Zahn J.-P.,  2012, Astronomy \& Astrophysics, 544, A132

\bibitem[\protect\citeauthoryear{Rieutord}{Rieutord}{2001}]{rieutord2001ekman}
Rieutord M.,  2001, The Astrophysical Journal, 550, 443

\bibitem[\protect\citeauthoryear{Rieutord}{Rieutord}{2004}]{rieutord2004evolution}
Rieutord M.,  2004, in Symposium-International Astronomical Union. pp 394--403

\bibitem[\protect\citeauthoryear{Rieutord}{Rieutord}{2006}]{rieutord2006dynamics}
Rieutord M.,  2006, Astronomy \& Astrophysics, 451, 1025

\bibitem[\protect\citeauthoryear{Rieutord \& Valdettaro}{Rieutord \&
  Valdettaro}{1997}]{rieutord1997inertial}
Rieutord M.,  Valdettaro L.,  1997, Journal of Fluid Mechanics, 341, 77

\bibitem[\protect\citeauthoryear{Rieutord \& Valdettaro}{Rieutord \&
  Valdettaro}{2010}]{rieutord2010viscous}
Rieutord M.,  Valdettaro L.,  2010, Journal of Fluid Mechanics, 643, 363

\bibitem[\protect\citeauthoryear{Rivinius, Townsend, Kochukhov, {\v{S}}tefl,
  Baade, Barrera  \& Szeifert}{Rivinius et~al.}{2012}]{rivinius2012basic}
Rivinius T.,  Townsend R. H.~D.,  Kochukhov O.,  {\v{S}}tefl S.,  Baade D.,
  Barrera L.,   Szeifert T.,  2012, Monthly Notices of the Royal Astronomical
  Society, 429, 177

\bibitem[\protect\citeauthoryear{Rivinius, Carciofi  \& Martayan}{Rivinius
  et~al.}{2013}]{rivinius2013classical}
Rivinius T.,  Carciofi A.~C.,   Martayan C.,  2013, The Astronomy and
  Astrophysics Review, 21, 69

\bibitem[\protect\citeauthoryear{Roberts}{Roberts}{1968}]{roberts1968thermal}
Roberts P.~H.,  1968, Philosophical Transactions of the Royal Society of London
  A: Mathematical, Physical and Engineering Sciences, 263, 93

\bibitem[\protect\citeauthoryear{Rogers, Lin, McElwaine  \& Lau}{Rogers
  et~al.}{2013}]{rogers2013internal}
Rogers T.~M.,  Lin D.,  McElwaine J.~N.,   Lau H. H.~B.,  2013, The
  Astrophysical Journal, 772, 21

\bibitem[\protect\citeauthoryear{Sana et~al.,}{Sana
  et~al.}{2012}]{sana2012binary}
Sana H.,  et~al., 2012, Science, 337, 444

\bibitem[\protect\citeauthoryear{Schaeffer}{Schaeffer}{2013}]{schaeffer2013efficient}
Schaeffer N.,  2013, Geochemistry, Geophysics, Geosystems, 14, 751

\bibitem[\protect\citeauthoryear{Schaeffer, Jault, Nataf  \&
  Fournier}{Schaeffer et~al.}{2017}]{schaeffer2017geodynamo}
Schaeffer N.,  Jault D.,  Nataf H.-C.,   Fournier A.,  2017, Geophysical
  Journal International, 211, 1

\bibitem[\protect\citeauthoryear{Schrinner, Petitdemange  \& Dormy}{Schrinner
  et~al.}{2012}]{schrinner2012dipole}
Schrinner M.,  Petitdemange L.,   Dormy E.,  2012, The Astrophysical Journal,
  752, 121

\bibitem[\protect\citeauthoryear{Seilmayer, Stefani, Gundrum, Weier, Gerbeth,
  Gellert  \& R{\"u}diger}{Seilmayer et~al.}{2012}]{seilmayer2012experimental}
Seilmayer M.,  Stefani F.,  Gundrum T.,  Weier T.,  Gerbeth G.,  Gellert M.,
  R{\"u}diger G.,  2012, Physical Review Letters, 108, 244501

\bibitem[\protect\citeauthoryear{Seshasayanan, Gallet  \&
  Alexakis}{Seshasayanan et~al.}{2017}]{seshasayanan2017transition}
Seshasayanan K.,  Gallet B.,   Alexakis A.,  2017, Physical Review Letters,
  119, 204503

\bibitem[\protect\citeauthoryear{Seyed-Mahmoud, Aldridge  \&
  Henderson}{Seyed-Mahmoud et~al.}{2004}]{seyed2004elliptical}
Seyed-Mahmoud B.,  Aldridge K.,   Henderson G.,  2004, Physics of the Earth and
  Planetary Interiors, 142, 257

\bibitem[\protect\citeauthoryear{Simitev \& Busse}{Simitev \&
  Busse}{2017}]{simitev2017baroclinically}
Simitev R.~D.,  Busse F.~H.,  2017, Geophysical \& Astrophysical Fluid
  Dynamics, 111, 369

\bibitem[\protect\citeauthoryear{Spiegel \& Veronis}{Spiegel \&
  Veronis}{1960}]{spiegel1960boussinesq}
Spiegel E.~A.,  Veronis G.,  1960, The Astrophysical Journal, 131, 442

\bibitem[\protect\citeauthoryear{Spruit}{Spruit}{1999}]{spruit1999differential}
Spruit H.~C.,  1999, Astronomy \& Astrophysics, 349, 189

\bibitem[\protect\citeauthoryear{Spruit}{Spruit}{2002}]{spruit2002dynamo}
Spruit H.~C.,  2002, Astronomy \& Astrophysics, 381, 923

\bibitem[\protect\citeauthoryear{Spruit \& Knobloch}{Spruit \&
  Knobloch}{1984}]{spruit1984baroclinic}
Spruit H.~C.,  Knobloch E.,  1984, Astronomy and Astrophysics, 132, 89

\bibitem[\protect\citeauthoryear{Stefani, Giesecke, Weber  \& Weier}{Stefani
  et~al.}{2016}]{stefani2016synchronized}
Stefani F.,  Giesecke A.,  Weber N.,   Weier T.,  2016, Solar Physics, 291,
  2197

\bibitem[\protect\citeauthoryear{Stello, Cantiello, Fuller, Huber, Garc{\'\i}a,
  Bedding, Bildsten  \& Aguirre}{Stello et~al.}{2016}]{stello2016prevalence}
Stello D.,  Cantiello M.,  Fuller J.,  Huber D.,  Garc{\'\i}a R.~A.,  Bedding
  T.~R.,  Bildsten L.,   Aguirre V.~S.,  2016, Nature, 529, 364

\bibitem[\protect\citeauthoryear{Strugarek, Beaudoin, Charbonneau, Brun  \& do
  Nascimento}{Strugarek et~al.}{2017}]{strugarek2017reconciling}
Strugarek A.,  Beaudoin P.,  Charbonneau P.,  Brun A.~S.,   do Nascimento
  J.-D.,  2017, Science, 357, 185

\bibitem[\protect\citeauthoryear{Su et~al.,}{Su et~al.}{2013}]{su2013asteroid}
Su K.~Y.,  et~al., 2013, The Astrophysical Journal, 763, 118

\bibitem[\protect\citeauthoryear{Szklarski \& Arlt}{Szklarski \&
  Arlt}{2013}]{szklarski2013nonlinear}
Szklarski J.,  Arlt R.,  2013, Astronomy \& Astrophysics, 550, A94

\bibitem[\protect\citeauthoryear{Tayler}{Tayler}{1973}]{tayler1973adiabatic}
Tayler R.,  1973, Monthly Notices of the Royal Astronomical Society, 161, 365

\bibitem[\protect\citeauthoryear{Tilgner}{Tilgner}{2005}]{tilgner2005precession}
Tilgner A.,  2005, Physics of Fluids, 17, 034104

\bibitem[\protect\citeauthoryear{Vantieghem, C{\'e}bron  \& Noir}{Vantieghem
  et~al.}{2015}]{vantieghem2015latitudinal}
Vantieghem S.,  C{\'e}bron D.,   Noir J.,  2015, Journal of Fluid Mechanics,
  771, 193

\bibitem[\protect\citeauthoryear{Vidal \& C\'ebron}{Vidal \&
  C\'ebron}{2017}]{vidal2017inviscid}
Vidal J.,  C\'ebron D.,  2017, \mn@doi [Journal of Fluid Mechanics]
  {10.1017/jfm.2017.689}, 833, 469

\bibitem[\protect\citeauthoryear{Weber, Galindo, Stefani  \& Weier}{Weber
  et~al.}{2015}]{weber2015tayler}
Weber N.,  Galindo V.,  Stefani F.,   Weier T.,  2015, New Journal of Physics,
  17, 113013

\bibitem[\protect\citeauthoryear{Wu \& Roberts}{Wu \&
  Roberts}{2009}]{wu2009dynamo}
Wu C.-C.,  Roberts P.~H.,  2009, Geophysical \& Astrophysical Fluid Dynamics,
  103, 467

\bibitem[\protect\citeauthoryear{Yadav \& Christensen}{Yadav \&
  Christensen}{2013}]{yadav2013scaling}
Yadav R. K.and~Gastine T.,  Christensen U.~R.,  2013, Icarus, 225, 185

\bibitem[\protect\citeauthoryear{Yadav, Gastine, Christensen  \& Duarte}{Yadav
  et~al.}{2013}]{yadav2013consistent}
Yadav R.~K.,  Gastine T.,  Christensen U.~R.,   Duarte L. D.~V.,  2013, The
  Astrophysical Journal, 774, 6

\bibitem[\protect\citeauthoryear{Zahn}{Zahn}{1966}]{zahn1966marees}
Zahn J.-P.,  1966, in Annales d'Astrophysique. p.~313

\bibitem[\protect\citeauthoryear{Zahn}{Zahn}{1992}]{zahn1992circulation}
Zahn J.-P.,  1992, Astronomy and Astrophysics, 265, 115

\bibitem[\protect\citeauthoryear{Zahn}{Zahn}{2008}]{zahn2008instabilities}
Zahn J.-P.,  2008, Proceedings of the International Astronomical Union, 4, 47

\bibitem[\protect\citeauthoryear{Zahn, Brun  \& Mathis}{Zahn
  et~al.}{2007}]{zahn2007magnetic}
Zahn J.-P.,  Brun A.~S.,   Mathis S.,  2007, Astronomy \& Astrophysics, 474,
  145

\bibitem[\protect\citeauthoryear{Zhang, Chan, Zou, Liao  \& Schubert}{Zhang
  et~al.}{2003}]{zhang2003three}
Zhang K.,  Chan K.~H.,  Zou J.,  Liao X.,   Schubert G.,  2003, The
  Astrophysical Journal, 596, 663

\bibitem[\protect\citeauthoryear{Zheng, Lin, Kouwenhoven, Mao  \& Zhang}{Zheng
  et~al.}{2017}]{zheng2017clearing}
Zheng X.,  Lin D. N.~C.,  Kouwenhoven M. B.~N.,  Mao S.,   Zhang X.,  2017, The
  Astrophysical Journal, 849, 98

\makeatother
\end{thebibliography}




\appendix

\section{Weakening of the tidal instability when $1 \lesssim N_0/\Omega_s \leq 2$}
\label{sec:append_weakening}
\begin{figure}
	\centering
	\includegraphics[width=0.45\textwidth]{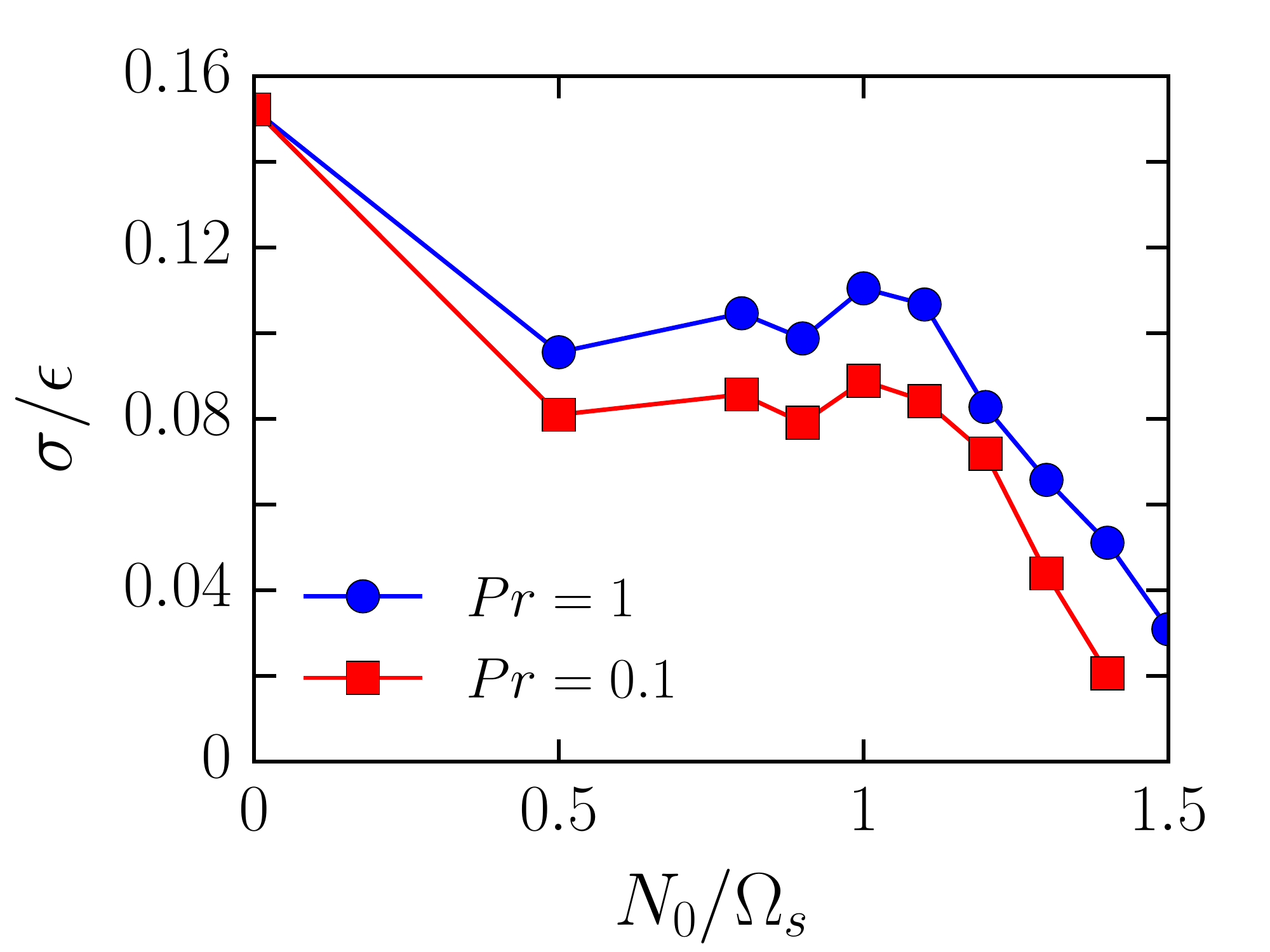}
	\caption{Normalised growth rate of the tidal instability $\sigma/\epsilon$ for varying $N_0/\Omega_s$. Simulations at $Ek=10^{-4}, \epsilon=0.2$, $Pr=1$ (circles) and $Pr=0.1$ (squares). }
	\label{fig:Fig_collapse_eps02_sigma}
\end{figure}

\begin{figure}
	\centering
	\includegraphics[width=0.45\textwidth]{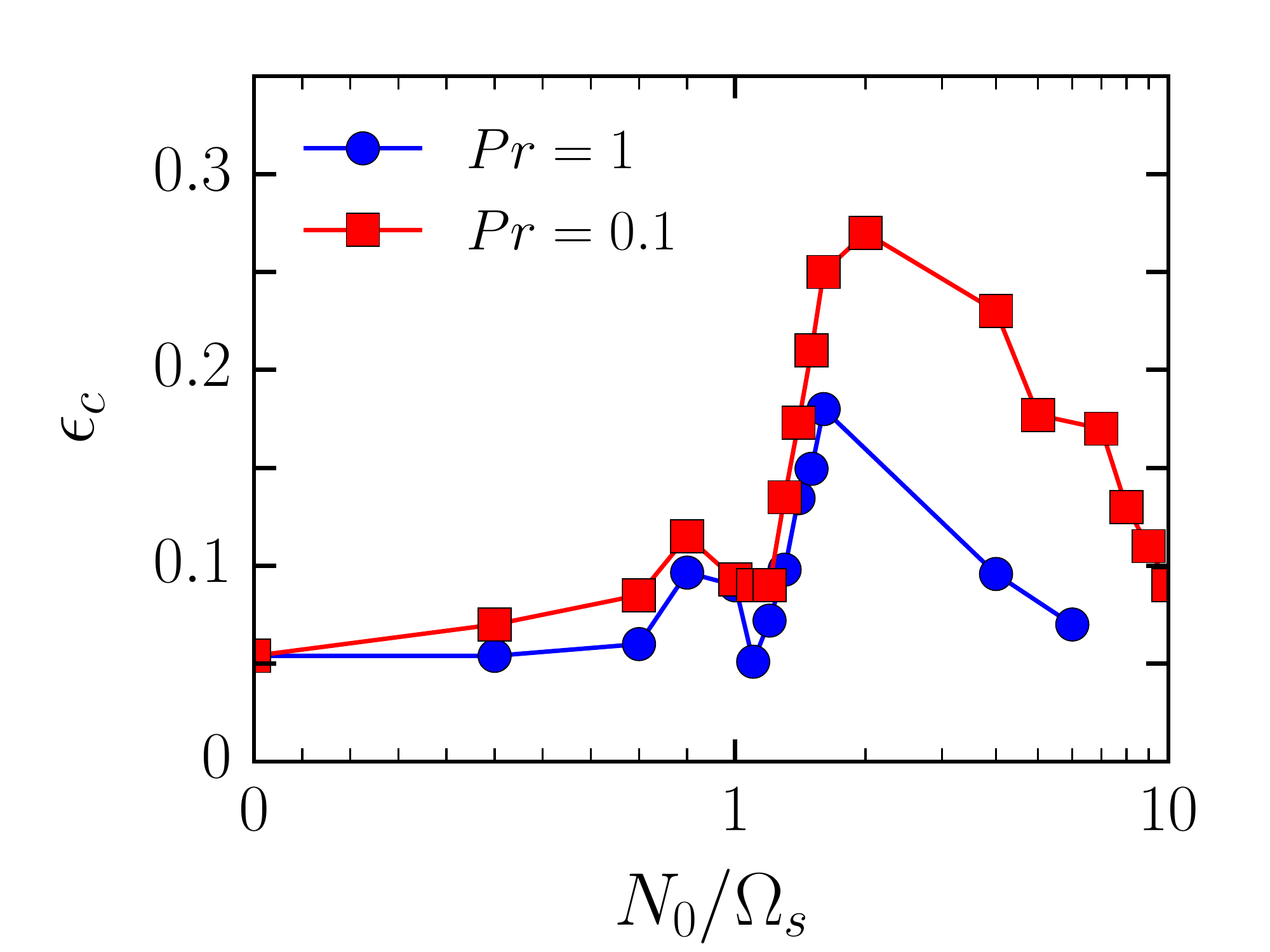}
	\caption{Threshold $\epsilon_c$ of the tidal instability for varying $N_0/\Omega_s$. Simulations at $Ek=10^{-4}$, $Pr=1$ (circles) and $Pr=0.1$ (squares). To determine $\epsilon_c$ we have performed simulations for several ellipticity $\epsilon$. Horizontal axis is linear between 0 and 1, then it is logarithmic.
	}
	\label{fig:Fig_collapse_eps02_epsca}
\end{figure}

The energy collapse of nonlinear flows in figure \ref{Fig_TDEIm_eps02} (b), responsible for the absence of mixing in figure \ref{Fig_Mixing} (b) when $1 \leq N_0/\Omega_s \leq 2$, is due to diffusive effects at the moderately small value $Ek=10^{-4}$ and $Pr=1$.
We performed simulations at $Ek=10^{-4}$ and $Pr=0.1$, i.e. for a thermal diffusion ten times larger than viscous diffusion.
In figure \ref{fig:Fig_collapse_eps02_sigma}, we show the normalised growth rate $\sigma/\epsilon_c$ for varying $N_0/\Omega_s$. 
When $N_0/\Omega_s \lesssim 1$ the growth rates for both $Pr=1$ and $Pr=0.1$ are weakly affected and almost insensitive to $N_0/\Omega_s$.
However for stronger stratifications, the growth rates are strongly reduced.
When $1.5 \leq N_0/\Omega_s \leq 2$, the tidal instability is even lost in simulations at $\epsilon=0.2$.
Thus, the critical ellipticity $\epsilon_c$ above which the tidal instability is triggered evolves with $N_0/\Omega_s$ at our moderate Ekman number.
To quantify this effect, we show in figure \ref{fig:Fig_collapse_eps02_epsca} how $\epsilon_c$ evolves as a function of $N_0/\Omega_s$.
In the range of interest $1 \leq N_0/\Omega_s \leq 2$, $\epsilon_c$ quickly increases with $N_0^2/\Omega_s^2$.
Hence, nonlinear curves in figures \ref{Fig_TDEIm_eps02} (b) and \ref{Fig_Mixing} (b) have not been obtained for a constant supercriticality $\epsilon/\epsilon_c$.
This phenomenon explains why the amplitude of nonlinear flows quickly drops for $1\leq N_0/\Omega_s \leq 2$,
because simulations at $N_0/\Omega_s \leq 1$ are about 4 times critical while the ones at $1\leq N_0/\Omega_s \leq 2$ are only barely supercritical.
Finally, for stronger stratification ($N_0/\Omega_s \gg 2$), the threshold $\epsilon_c$ decreases back to values close to the ones without stratification.
This is the reason why we observe the onset of the tidal instability for these stratifications in figure \ref{Fig_TDEIm_eps02}.
The more $N_0/\Omega_s$ increases, the more radial motions are inhibited and become of short wavelength in the linear growth of the instability and toroidal motions are favoured. The latter motions are the least diffusively damped flows with stress-free boundary conditions \citep[e.g.][]{rieutord2001ekman}. Hence, the combined effects of diffusion and stronger stratification favour toroidal motions and decrease the threshold of the tidal instability.


\bsp	
\label{lastpage}
\end{document}